\documentclass[preprint,prb,groupedaddress,showpacs,amsmath,amssymb,a4paper]{revtex4}

\usepackage{graphicx}
\usepackage{dcolumn}
\usepackage{bm}

\newcommand{\arccosh}{\rm arccosh}

\begin{document}
\title{Nonlocal electrodynamics of long ultra-narrow Josephson
junctions: Experiment and theory}

\author{A.~A.~Abdumalikov~Jr.$^\dag$}
\email{abdumalikov@physik.uni-erlangen.de}
\author{V.~V.~Kurin$^\ddag$}
\author{C.~Helm$^\S$}
\author{A.~De~Col$^\S$}
\author{Y.~Koval$^\dag$}
\author{A.~V.~Ustinov$^\dag$}
\affiliation{\dag Physikalisches Institut III, Universit\"at
Erlangen-N\"urnberg, Erlangen D-91058, Germany \\
\ddag Institute for Physics of Microstructure (RAS), Nizhniy
Novgorod, Russia \\
\S Institut f\"ur Theoretische Physik, ETH H{\"o}nggerberg,
Z{\"u}rich 8093, Switzerland}

\date{\today}

\begin{abstract}
We experimentally and theoretically investigate electromagnetic
cavity modes in ultra-narrow Al-AlO$_x$-Al and Nb-AlO$_x$-Nb long
Josephson junctions. Experiments show that the voltage spacing
between  the Fiske steps on the current-voltage characteristics of
sub-$\mu$m wide and several hundred $\mu$m long Al-AlO$_x$-Al and
Nb-AlO$_x$-Nb Josephson junctions increases when decreasing the
width of a junction. This effect is explained by stray magnetic
fields, which become important for narrow junctions. Theoretical
estimates of the Fiske step voltage based on a nonlocal wave
propagation equation are in good agreement with our experimental
data. Using the nonlocal model, we determine the size and mass of
a Josephson vortex by means of a variational approach, and relate
vortex size to the experimentally measured critical magnetic field
of the junction.
\end{abstract} 

\maketitle

\section{\label{intro}Introduction}
\begin{figure}
\includegraphics{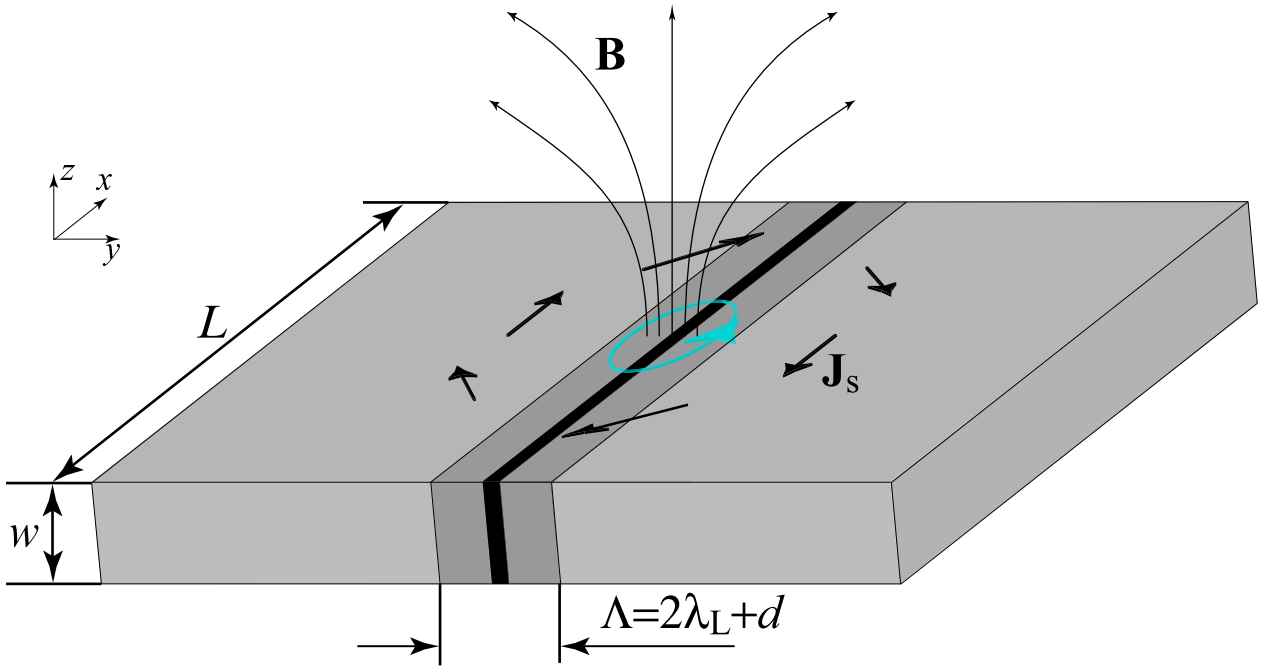}
\caption{Schematic view of an edge-type long Josephson junction of
   width $w$ and length $L$. The lines show the magnetic field
   $\mathbf B$, due to vortex, which penetrates the electrodes and
   extends into free space. The extended stray magnetic field creates
   the additional surface current $\mathbf J_S$.}
  \label{niceartistic}
\end{figure}

The electromagnetic properties of long Josephson tunnel junctions
have been the subject of intensive studies over the past four
decades \cite{Yanson,BarPat,Likharev,Ustinov}. Most investigations
of junction dynamics have been based on local theory. This theory
assumes that electromagnetic fields are concentrated in the
junction barrier, and neglects the distribution of the magnetic
and electric fields outside the junction. However, in reality the
fields penetrate the superconducting electrodes and extend into
free space (Fig.~\ref{niceartistic}). Local theory breaks down
when the fields outside the junction contribute significantly to
the junction energy.

The magnetic field inside the superconducting electrodes of a
Josephson junction becomes important when its characteristic decay
length is of the order of (or larger then) the characteristic
spatial scale of variation of the Josephson phase $\varphi$. This
type of nonlocality we call \emph{internal}, because it depends
only on the field distribution inside the junction and its
electrodes, and neglects extended, stray, fields outside the
junction.

When the energy of stray fields becomes comparable to the energy
of the field inside the junction, one has to take stray fields
into account. These stray fields generate an additional surface
current $\mathbf J_S$, which can be represented as a functional of
the magnetic field distribution inside the junction, leading to a
nonlocal equation. Because these effects depend on the geometry of
the junction, we call them \emph{geometrical nonlocal} effects.

Detailed knowledge of Josephson vortex dynamics is important for
understanding the magnetic flux motion and related phenomena in
superconductors. In a long junction, a vortex behaves like a
classical particle, which is characterized by the spatial
coordinate of its center and the effective mass. The interest to
experimentally verify theoretically predicted nonlocal corrections
to the conventional model has been biased by recent studies of
long ultra-narrow junctions where Josephson vortices behave as
macroscopic quantum objects\cite{ustinovvortex, Fistul}. In very
narrow junctions, the \emph{effective dynamical mass} of the
vortex is expected to be influenced by nonlocally described stray
magnetic fields. Most of the theoretical papers assume, however,
that the effective dynamical mass of a Josephson vortex is
proportional to the width of the junction as expected in the local
theory \cite{Hermon, Kato,Shnirman}.

In this paper, we experimentally verify that, in very narrow long
Josephson junctions, one has to indeed take into account the
vortex mass correction due to the stray fields outside the
junction. Significant deviations from the local theory have
previously been observed in the current-voltage characteristics of
long junctions of small width\cite{Koval}. We systematically study
these effects for different junction materials, and compare the
experimental findings with the nonlocal sine-Gordon model.

A Josephson junction can be considered quasi-one-dimensional, as
long as its transverse dimension, the width $w$ in $z$-direction
(see Fig.~\ref{niceartistic}), is smaller than the Josephson
penetration depth
\begin{equation}\label{Josephsonpenetration}
\lambda_J=\sqrt{\frac{\Phi_0}{2\pi\mu_0 j_c\Lambda}},
\end{equation}
where $\Phi_0$ is the magnetic flux quantum, $j_c$ is the critical
current density across the tunnel barrier, $\Lambda=2\lambda_L+d$
is the magnetic thickness of the junction, $\lambda_L$ is the
London penetration depth and $d$ is the barrier thickness
(typically, $d\ll\lambda_L$). In this limit ($w\ll\lambda_J$), the
long junction is described by the perturbed one-dimensional
sine-Gordon equation for the Josephson phase difference $\varphi$

\begin{equation}\label{SGE}
    \lambda_J^2\varphi_{xx}-\omega_p^{-2}\varphi_{tt}-\omega_p^{-1}\alpha\varphi_t-\sin\varphi=\gamma,
\end{equation}

\noindent where subscripts mean partial derivatives with respect
to the spatial coordinate $x$ along the junction and time $t$. The
parameter
\begin{equation}
\omega_p=\sqrt{\frac{2\pi j_c}{\Phi_0C}}
\end{equation}
is the Josephson plasma frequency, $C$ is the specific capacitance
of the junction per unit area, and $\alpha$ is the dissipation
coefficient due to the quasiparticle tunnelling across the
barrier. The bias current density $\gamma=j/j_c$ is normalized to
the critical current density $j_c$  and is, in general, dependent
on $x$.

Equation~(\ref{SGE}) is derived using the following assumptions:
i) $\lambda_L\ll w \ll \lambda_J$ and ii) $\hbar\omega\ll\Delta$,
where $\omega$ is the oscillation frequency of the phase
$\varphi$, and $\Delta$ is the superconducting energy gap of the
electrodes. The breakdown of any of these conditions requires
modification of Eq.~(\ref{SGE}). As mentioned above, for large
London penetration depths $\lambda_L\geq\lambda_J$ (or small
Josephson penetration depths $\lambda_J$) the magnetic field
inside the superconductor starts to play an important role. On the
other hand, for sufficiently small junction width
$w\sim\lambda_L$, the stray fields outside the junction have to be
taken into account. In either case, the dynamics of the phase
$\varphi$ has to be described by an integro-differential equation,
i.e. the problem becomes nonlocal. When the oscillation frequency
$\omega$ is comparable to the superconducting gap $\Delta$, the
frequency dependence of the London penetration depth $\lambda_L
(\omega)$ according to the microscopic theory ({\em material
dispersion}) should also be taken into account
\cite{Mattis,Lee,Tinkham}.

The nonlocal equation describing Josephson phase dynamics can be
written in the general form
\begin{equation}\label{NSGE}
\lambda_J^2\frac{\partial} {\partial x}\int
dx'Q(x,x')\varphi_{x'}(x')-
\omega_p^{-2}\varphi_{tt}-\omega_p^{-1}\alpha\varphi_t-\sin\varphi=\gamma,
\end{equation}
where the function $Q(x,x')$ is the nonlocality kernel, which
needs to be specifically determined for each nonlocal problem. In
the local case $Q(x,x')=\delta(x-x')$, where $\delta(x)$ is a
delta function. Several theoretical approaches have been proposed
for different types of nonlocality, in the case of an infinitely
long Josephson junction. In the limit of an infinitely long
junction, the kernel of the integral in Eq.~(\ref{NSGE}) is
reduced to the function of a single argument, $Q(x,x')=Q(|x-x'|)$.

\begin{figure}[hbt]
  \includegraphics[scale=0.60]{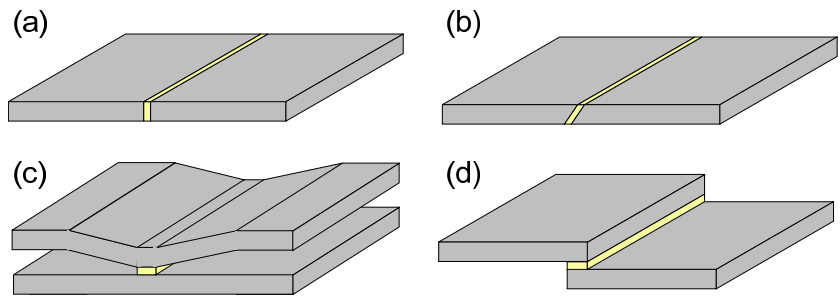}
  \caption{Different geometries of long Josephson junctions formed between two
  superconducting films: (a) edge-type junction;
  (b) ramp-type junction; (c) window junction; (d) overlap
  junction.}
  \label{allgeom}
\end{figure}

Nonlocal models can be divided into two groups, those treating
internal nonlocality inside bulk junctions \cite{Gur, AliSil}, and
those dealing with nonlocal effects due to outer stray fields
resulting from the geometry of the junction and its electrodes
\cite{Lapir, Kupriyanov, IvanSob, Kogansemi, Caputo2, Ivan,
MinSnap, MinSnapcheren, LomKuz, Lomtev, Caputo1, Kurin,
Kogan,groenbech}. The latter group of theories takes into account
the field configuration not only inside the junction, but also
around it, and incorporates the finite size of the sample and its
shape-dependent magnetic properties. The shorter the junction is
in $x$-direction (see Fig.~\ref{niceartistic}), or the narrower it
is in $z$- direction, the larger the effect of the geometrical
nonlocality. Four typical geometries of long Josephson junctions
are presented in Fig.~\ref{allgeom}. These are (a) edge-type, (b)
ramp-type, (c) window and (d) overlap long junctions. In the ideal
case of an infinitely long junction, the electrodynamic problem
posed by (a)--(c) type junctions have been solved. In particular,
theoretical models for the geometrical nonlocality have analyzed
the following cases: 1) edge-type junctions between thick films
$w\gtrsim\lambda_L$, neglecting internal nonlocal effects
\cite{Ivan,groenbech}; 2) edge-type junctions between thin films
$w<\lambda_L$ (Pearl's limit) \cite{Pearl,IvanSob,Kogan}; 3)
edge-type and 4) ramp type junctions between films of arbitrary
width, taking into account both the internal and external nonlocal
problems \cite{LomKuz,Lomtev}; 5) a variable thickness bridge
above a ground plane \cite{Lapir, Kupriyanov} and 6) window
junctions \cite{Caputo2, Caputo1, Kurin}.

In this paper, we study the effect of geometrical nonlocality by
analyzing experiments on extremely narrow long Josephson
junctions, down to a width $w$ of $0.1~\mu$m. The paper is
organized as follows. In Sec.~\ref{fabricsec} we describe the
geometries and preparation process of ultra-narrow low-$T_c$ long
Josephson junctions. In Sec.~\ref{exp} we present experimental
results obtained for Nb-AlO$_x$-Nb and Al-AlO$_x$-Al junctions.
Theoretical models based on nonlocal electrodynamics, and their
comparison with experiment, are presented in Sec.~\ref{theory}.
Section~\ref{conclusion} contains concluding remarks.

\section{\label{fabricsec}Fabrication and geometry of ultra-narrow
long Josephson junctions}

The fabrication process of high quality long Josephson junctions
in order to study the effects of nonlocal electrodynamics should
satisfy the following three requirements: (i) ability to vary the
junction width, from several micrometers down to several hundred
nanometers, (ii) constant width along the junction length
$L\gg\lambda_J$, which is typically several hundred micrometers,
and (iii) an idle area of overlapping electrodes (window) that is
as small as possible, so as to prevent it influencing wave
propagation in the junction. We prepared samples using two
different technologies based on aluminium and niobium. All samples
were fabricated on a thermally oxidized Si substrates.

\subsection{Al-AlO$_x$-Al junctions}

\begin{figure}
  \includegraphics[scale=0.75]{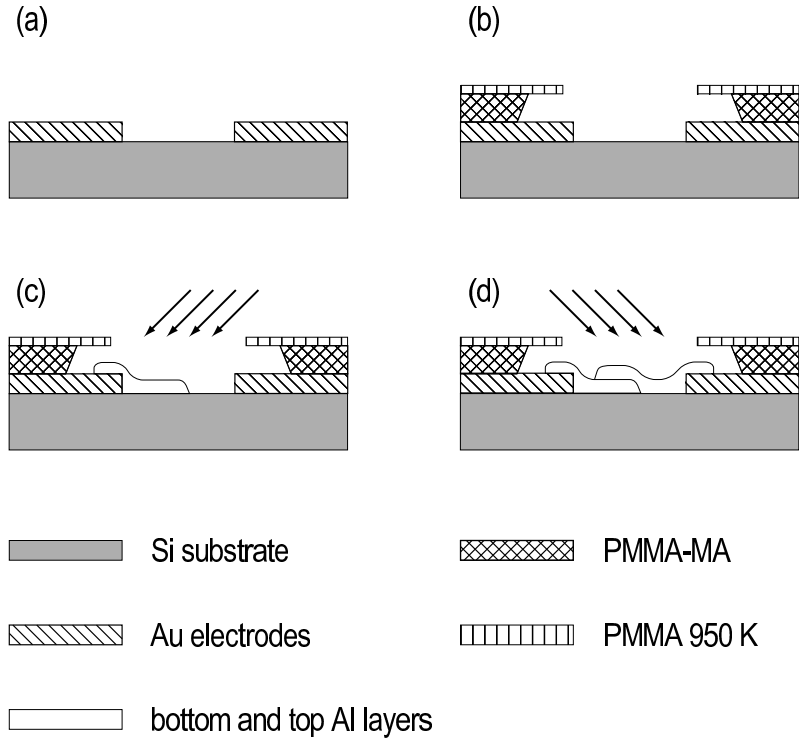}
  \caption{Schematic diagram of the fabrication procedure for Al
  junctions. (a) Deposition of Au electrodes; (b) formation
  of a window in the PMMA resist for shadow evaporation; (c,d)
  deposition of top and bottom Al electrodes.}
  \label{ALALAL}
\end{figure}

The Al-AlO$_x$-Al junctions were prepared with the shadow
evaporation technique. This method for the preparation of
sub-$\mu$m Al-AlO$_x$-Al tunnel junctions is well known (see, e.g.
Ref.~\onlinecite{Dolan}). However, we cannot apply this method
directly. A hanging bridge of electron resist (e.g. PMMA) or
another material is usually used for shadowing during the
evaporation of Al. As the length of the bridge becomes longer the
bridge starts to sag. Thus, for long junctions (several hundred
micrometers) the bridge cannot be well fixed. Therefore, in our
preparation method, we use a \emph{"shadowing window"}.
Schematically the fabrication steps are shown in
Fig.~\ref{ALALAL}.

\begin{figure}
  \includegraphics[scale=1.1]{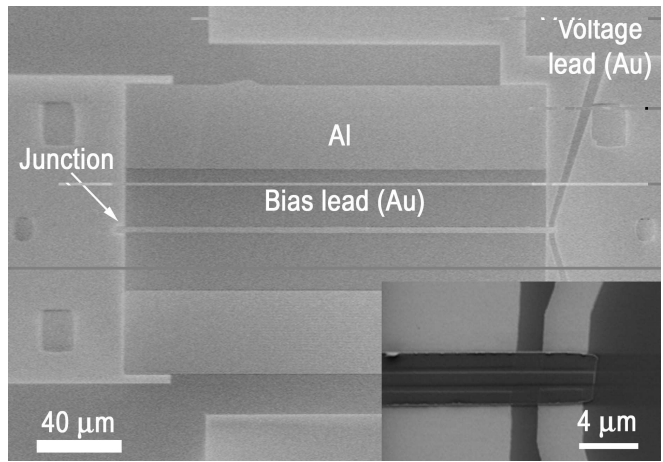}
  \caption{SEM picture of Al-AlO$_x$-Al junction. The inset
  shows an enlarged view of the connection of current and potential electrodes to
  the junction.}
  \label{SEMpicture}
\end{figure}

The first step was the formation of Au electrodes, used for
spatially-uniform bias current injection and voltage measurement
leads (Fig.~\ref{ALALAL}\,a). These electrodes were formed with
the help of electron-beam lithography, thermal evaporation and the
lift-off technique. The Cr layer under the Au layer improves Au
adhesion to the substrate. The thicknesses of Cr and Au layers
were 10 nm and 60 nm, respectively. The electrodes for current
injection and voltage measurement across the junction have
different geometries. One sees in Fig.~\ref{SEMpicture} that the
bias current leads have a length almost equal to that of the
Josephson junction. The potential leads are much smaller, and are
connected to the junction only at the edges of the Al electrodes.
This geometry provides homogeneous current injection into long
Josephson junctions and allows four point measurements of junction
characteristics.

To form long Josephson junction by shadow evaporation, we used
double layer resist PMMA-MA/PMMA 950 K, which was spanned onto the
substrate with prepared electrodes. Then, using electron-beam
lithography, we opened a window between the Au electrodes, as
shown in Fig.~\ref{ALALAL}\,b. In the next step, we evaporated Al
onto a tilted substrate (see Fig.~\ref{ALALAL}\,c). The thickness
of this bottom Al layer was about 50 nm. Immediately after Al
evaporation O$_2$ was injected into the vacuum chamber. The
pressure of O$_2$ was kept at 10$^{-3}$ mBar for 6 minutes. Then,
a 200~nm thick top layer of Al was evaporated onto the tilted
substrate (Fig.~\ref{ALALAL}\,d). The overlap of the bottom and
top Al electrodes defines the junction area. A scanning electron
micrograph of the resulting Josephson junction and electrodes is
shown in Fig.~\ref{SEMpicture}.

\begin{figure}
  \includegraphics[scale=0.75]{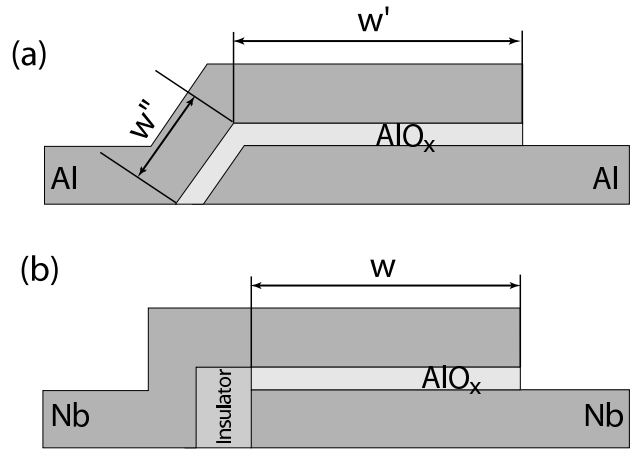}
  \caption{Schematic cross sections of experimentally studied
  Al-AlO$_x$-Al (a) and Nb-AlO$_x$-Nb (b) junctions.}
  \label{cross}
\end{figure}

Schematically, the cross section of the fabricated Josephson
junction is shown in Fig.~\ref{cross}\,a. Note that the total
junction width is composed of two sections, one, of width $w'$,
parallel to the film surface plane and the other, of width $w''$,
which is tilted. The total width is therefore $w=w'+w''$.
Typically the relation $w''\ll w'$ is well fulfilled. However, in
the case of the most narrow junctions, the contribution of $w''$
to the junction width becomes significant. We estimate
$w''\simeq60$ nm.

\subsection{Nb-AlO$_x$-Nb junctions}

\begin{figure}
  \includegraphics[scale=0.75]{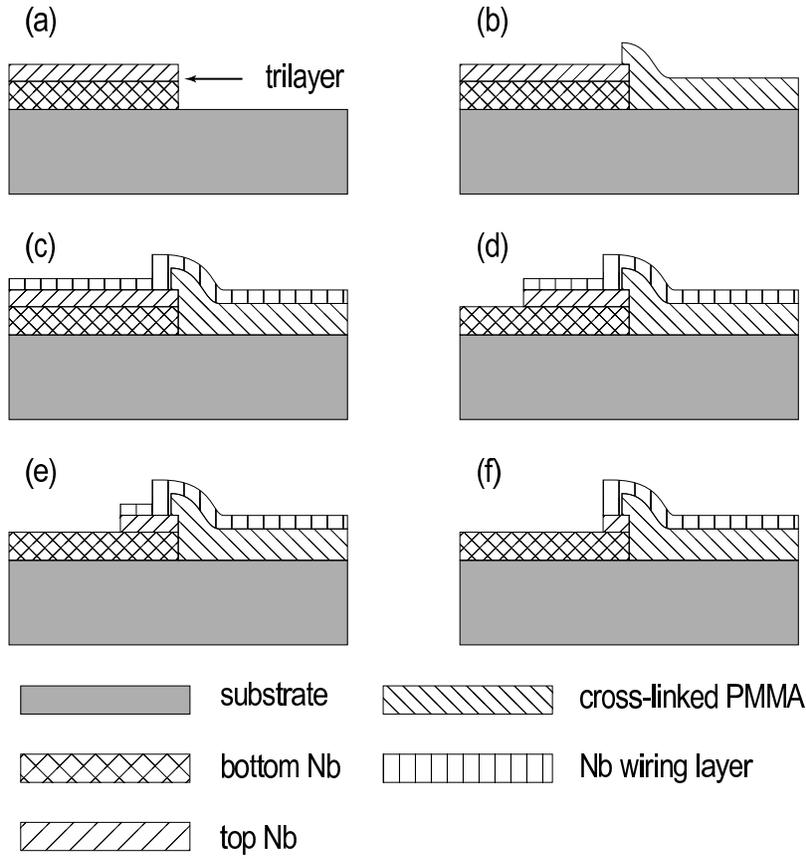}
  \caption{Schematic diagram of the fabrication procedure for Nb
  junctions. (a) Formation of the trilayer; (b) insulation of the
  trilayer edge; (c) deposition of Nb wiring; (d) removal of
  Nb wiring and top Nb from all areas except that of the
  junction, (e,f) narrowing of the junction using electron beam
  lithography and reactive ion etching (repeated several times).}
  \label{NBALNB}
\end{figure}

There are several different approaches for the preparation of
Josephson junctions using Nb-AlO$_x$-Nb trilayer technology. They
differ mainly in patterning and insulation of the junction.
Junctions with dimensions less than 1~$\mu$m usually have a window
geometry (see Fig.~\ref{allgeom}\,c), and are produced by reactive
ion etching and vapor deposition of a dielectric layer. Using this
approach, one could prepare sub $\mu$m junctions, however they
would not satisfy the requirement of small idle region between the
superconducting electrodes around the junction. In our group, we
developed an alternative method, which allows junctions on the
very edge of the Nb-AlO$_x$-Nb trilayer \cite{Koval}.
Schematically, this process is shown in Fig.~\ref{NBALNB}. The
junction cross section prepared in this way is shown in
Fig.~\ref{cross}\,b.

In the first step, the Nb-AlO$_x$-Nb trilayer region is defined by
lithography and reactive ion etching (Fig.~\ref{NBALNB}~a). The
next step is the fine definition of one of the edges of the
structure using electron beam lithography and etching. This
particular edge was insulated by cross-linked PMMA (see
Fig.~\ref{NBALNB}~b). The details of the cross-linking procedure
and the self-alignment of cross-linked PMMA along the trilayer
edge can be found in Ref.~\onlinecite{Koval}. In the next step,
the Nb wiring layer was deposited as shown in Fig.~\ref{NBALNB}~c.
Electron-beam lithography and reactive ion etching were used to
define the junction area (see Fig.~\ref{NBALNB}~d). After
performing a full set of measurements on the junction, the width
of the junction was iteratively decreased using electron beam
lithography and reactive ion etching, as shown in
Fig.~\ref{NBALNB}~e and f.

\section{\label{exp}Experiment}

In this section, we present the data acquired for Al-AlO$_x$-Al
and Nb-AlO$_x$-Nb ultra-narrow long Josephson junctions. A
magnetic field $H$ of up to $40\ $Oe was applied in the plane of
the junction, perpendicular to its long dimension. The magnetic
field of the Earth was screened out with a cryoperm shield. The
experimentally determined parameters of the junctions are
summarized in Table~\ref{tablePar}.
\begin{table}
  \caption{Parameters of the measured junctions.}
  \begin{tabular}{|c|c|c|c|c|c|}
  \hline
  junction material & $\lambda_L$, nm & $ J_c$, A/cm$^2$ & $\lambda_J$, $\mu$m & $\bar c$, m/s, & $V_{\rm gap}$, mV \\
  \hline Al ($T=0.3$ K)& 15 & 270 & 56 & $8\cdot10^6$ &0.35 \\
  Al ($T=0.9$ K)& 17 & 160 & 68 & -- &0.29 \\
  Nb ($T=4.2$ K)& 90 & 210 & 25 & $7 \cdot10^6$ & 2.7 \\
  \hline \end{tabular}\label{tablePar}
\end{table}
The London penetration depths for Al and Nb were found from period
$\Delta H$ of $I_c(H)$ at high fields for widest junctions, as
$\lambda_L=\Phi_0/(2 L \Delta H)$. The Josephson penetration depth
was found from the critical current density using
Eq.~(\ref{Josephsonpenetration}). The Swihart velocity was found
by fitting Fiske step voltages to the theoretical dependence (see
Sec.~\ref{Fisketheory}). The junction width was measured using a
scanning electron microscope.

In the following, we describe two main effects observed in our
ultra-narrow Josephson junctions, when the junction width $w$ is
decreased. These are 1) a reduction in the first critical field
$H_{c1}$, and 2) an increase of the Fiske step voltage spacing
$\Delta V_{FS}$. According to standard (local) theory, the first
critical field $H_{c1}$ and the Fiske step voltage should not
depend on the junction width \cite{BarPat}.

The first critical field corresponds to the field when a single
vortex enters the junction. In the local theory, the first
critical field $H_{c1}$ is inversely proportional to the Josephson
penetration depth $\lambda_J$. In the following sections we
present the experimental data for $H_{c1}$ in terms of an
effective length scale
\begin{equation}\label{lambdatilde}
{\tilde \lambda}_J =\frac{\Phi_0}{ \pi H_{c1}(2\lambda_L+t)}.
\end{equation}
Note that this equation is similar to the relation between
$\lambda_J$ and $H_{c1}$ in the local theory.

Fiske steps arise when the Josephson oscillation frequency
resonates with cavity modes of the junction. The voltages of these
steps according to local theory are given by
\begin{equation}\label{Fiskelocal}
V_n = \frac{\Phi_0}{2\pi} \frac{n\pi}{L}
\bar{c}=\frac{\Phi_0}{2\pi} \frac{n\pi}{L} \omega_p\lambda_J,
\end{equation}
where $\bar{c}$ is the Swihart velocity and $n$ is the step
number. Note that the Fiske step voltages $V_n$ are also
determined by the length scale $\lambda_J$, provided Josephson
plasma frequency does not change.

\subsection{Al-AlO$_x$-Al junctions}
\begin{figure}
  \includegraphics[scale=1.26]{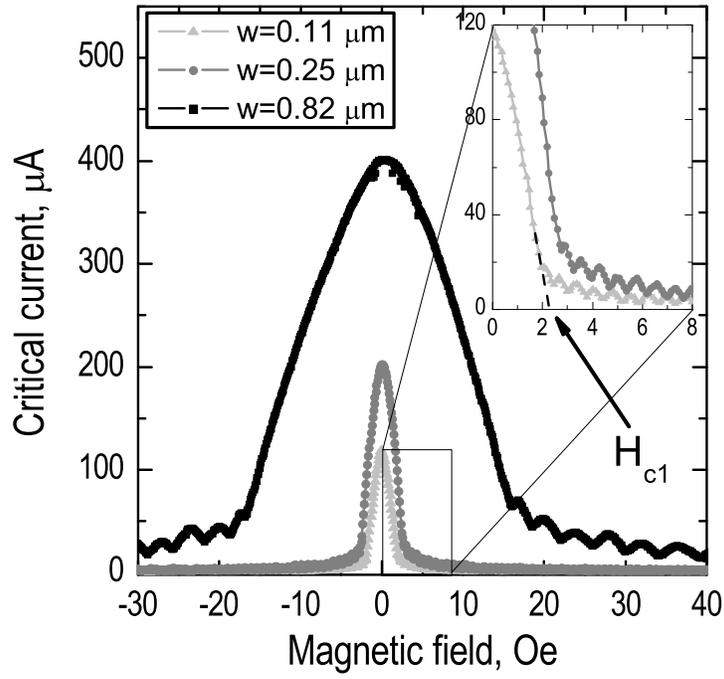}
  \caption{Critical current modulation $I_c(H)$ in magnetic field
  for long Al-AlO$_x$-Al Josephson junctions at $T=0.9$ K. The inset
  zooms in on the modulation patterns of the junctions of width
  $w=0.11~\mu$m and $w=0.25~\mu$m.}
  \label{Ich_Al}
\end{figure}

We measured a series of six $230\ \mu$m long Al-AlO$_x$-Al
junctions. The width $w$ of the junctions varied between $0.1\
\mu$m and $1\ \mu$m. The homogeneity of the junctions was verified
by measuring the modulation pattern of the critical current $I_c$
vs. field $H$ at $T=0.9$ K. In Fig.~\ref{Ich_Al}, $I_c(H)$
patterns of three samples of width $w=0.11,\ 0.25$ and $0.82\
\mu$m are presented. We find the first critical field by linearly
extrapolating the central peak in the critical current modulation
pattern (see inset of Fig.~\ref{Ich_Al}). We observe that the
first critical field $H_{c1}$ decreases with decreasing the
junction width $w$.

\begin{figure}
  \includegraphics[scale=1.26]{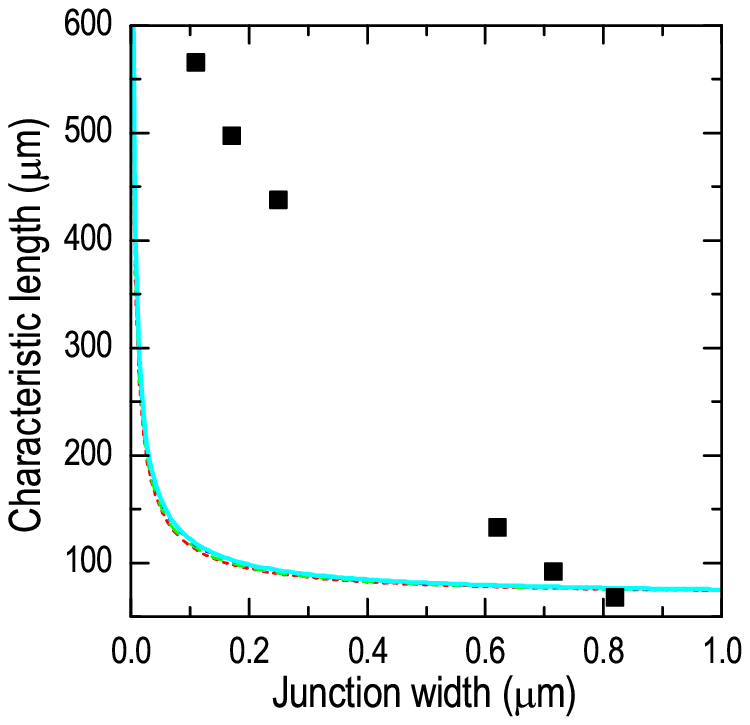}
  \caption{Characteristic spatial scale of the Josephson phase in
    Al-AlO$_x$-Al junctions. Squares indicate the characteristic
    length scale ${\tilde\lambda}_J$, as extracted from the measured
    critical field $H_{c1}$ at temperature $T=0.9$ K, using
    to Eq.~(\ref{lambdatilde}). Lines show
    theoretical estimations of the size $\tilde\lambda_J$ of a vortex,
    for details see Sec.~\ref{vortexsize}.}
  \label{lengthAL}
\end{figure}

The experimentally determined effective length scale
$\tilde\lambda_J$ is presented in Fig.~\ref{lengthAL} as black
squares. This length scale depends strongly on the junction width.
Theoretical curves are discussed below, in Sec.~\ref{vortexsize}.

\begin{figure}
  \includegraphics[scale=1.26]{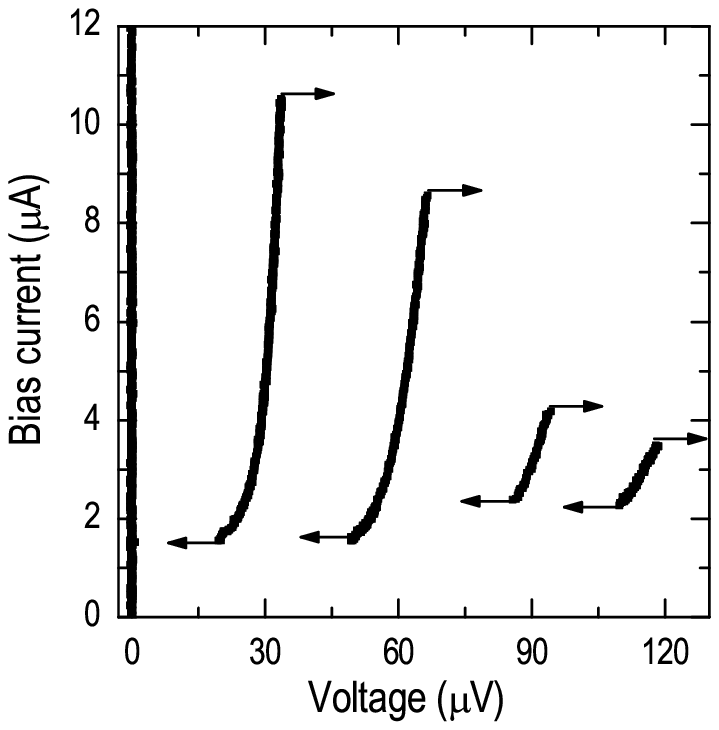}
  \caption{Superimposed traces of Fiske resonances of the
  current-voltage characteristics of Al-AlO$_x$-Al long
  Josephson junctions at several different magnetic fields. The
  junction width is $w=0.82\ \mu$m and the temperature is $T=0.3$ K.}
  \label{Fiskestep}
\end{figure}

Fiske steps of all junctions were measured by tracing the
current-voltage characteristics during a continuous sweep of the
external field $H$ in the range from $-40\ $Oe to $+40\ $Oe. These
measurements were performed at temperature $T=0.3$ K, because at
the higher temperature $T=0.9$ K high damping suppresses the Fiske
resonances. At $T=0.3$ K the first critical field $H_{c1}$ of the
wide junctions ($w\simeq 0.6-0.8~\mu$m) was larger than the
maximum field we could apply. In Fig.~\ref{Fiskestep}, the
superimposed characteristics are plotted for the junction of width
$w=0.82\ \mu$m.

\begin{figure}
  \includegraphics[scale=1.26]{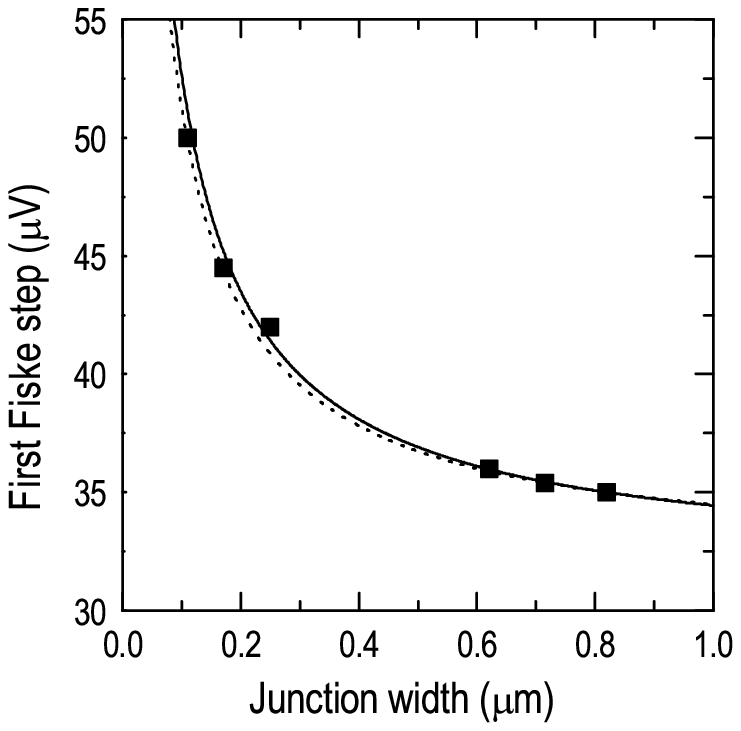}
  \caption{The width dependence of the first Fiske step voltage
  $V_{1}$ for Al-AlO$_x$-Al junctions. Symbols correspond to experimental
  data obtained at $T=0.3K$. The solid line corresponds to the theory of
  case \cite{Ivan} (c) and the dashed line to case \cite{LomKuz} (d).}
  \label{FFS_AL}
\end{figure}

We have observed that the positions of the Fiske steps strongly
depend on the width of the junction. This dependence for the first
step is shown in Fig.~\ref{FFS_AL} and compared with the theory
presented below in Sec.~\ref{Fisketheory}.

\subsection{Nb-AlO$_x$-Nb junctions}

\begin{figure}
  \includegraphics[scale=1.26]{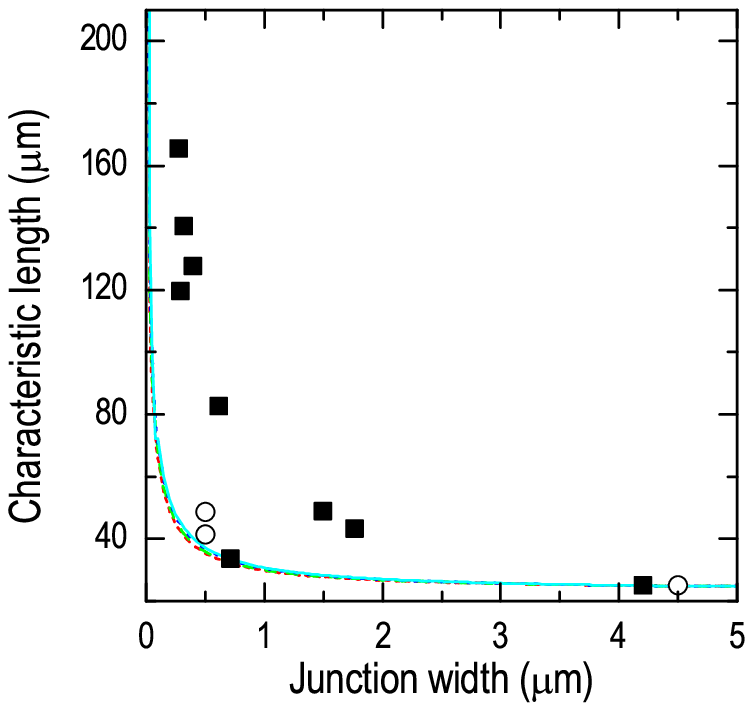}
  \caption{Characteristic spatial scale of the Josephson phase in
  Nb-AlO$_x$-Nb junction at $T=4.2$ K. The open circles denote the length
  scale ${\tilde \lambda}_J$ extracted from the experimental critical
  field $H_{c1}$ according to Eq.~(\ref{lambdatilde}) for new junctions.
  The solid symbols correspond to experimental data for $H_{c1}$ of Ref.~\onlinecite{Koval}.
  Lines correspond to the theoretical estimations of the size
  $\tilde \lambda_J$ of a vortex (see Sec.~\ref{vortexsize}).}
  \label{length}
\end{figure}

\begin{figure}
  \includegraphics[scale=1.26]{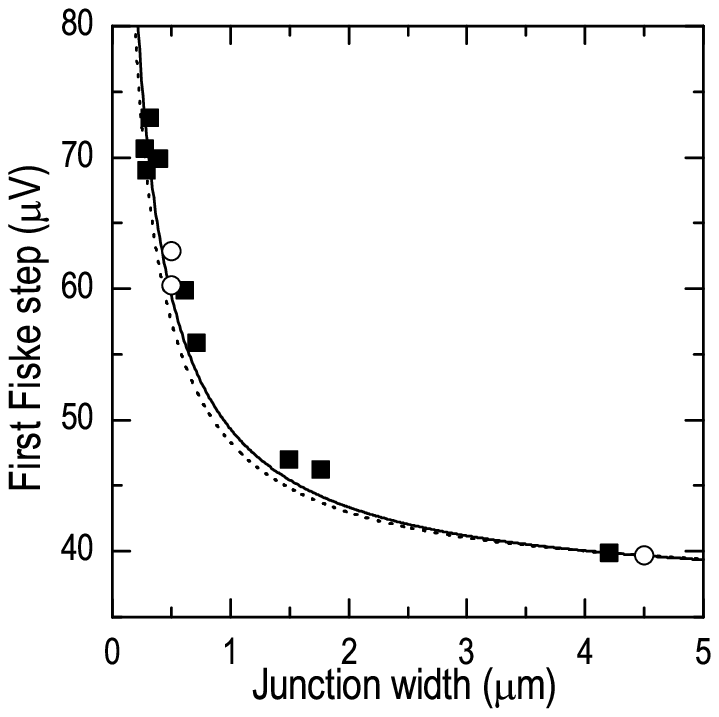}
  \caption{The width dependence of the maximum of the first Fiske step
  voltage $V_{1}$ for Nb-AlO$_x$-Nb junctions. The open symbol correspond
  to new experimental data and the data for solid symbols are taken from
  Ref.~\onlinecite{Koval}. The solid line is calculated
  using the theory of Ref.~\onlinecite{Ivan} and the dashed line
  corresponds to the theory of Ref.~\onlinecite{LomKuz}.}
  \label{FFS_NB}
\end{figure}

Results of systematic measurements of ten long narrow
Nb-AlO$_x$-Nb junctions were presented earlier in Ref.\
\onlinecite{Koval}. We have now additionally measured three more
junctions. Measurements were performed at temperature $T=4.2$ K.
In the present paper we analyze earlier data and our new results
more extensively. The junction length is $200\ \mu$m and the width
$w$ ranges from $0.3\ \mu$m to $4.5\ \mu$m. The Josephson
penetration depth was estimated from the critical current density
using Eq.~(\ref{Josephsonpenetration}) to be about $\lambda_J
\approx 25\ \mu$m. The critical current vs. magnetic field
patterns, reported in Ref. \onlinecite{Koval}, and those of the
new junctions show that all junctions are rather homogeneous. As
in the case of the Al-AlO$_x$-Al junctions, a decrease in the
first critical field $H_{c1}$ (Fig.~\ref{length}), and an increase
of the Fiske step voltage spacing $\Delta V_{\rm FS}$
(Fig.~\ref{FFS_NB}), with decreasing width $w$ of the junction
were observed in the Nb-AlO$_x$-Nb junctions.

\section{\label{theory}Comparison with theory}

In this section, we present a short review of the existing
theoretical models for the nonlocal effects in edge-type Josephson
junctions (Fig.~\ref{niceartistic}). In particular, we are
interested in the nonlocal kernels $Q(x,x^\prime)$ to be used in
Eq.~(\ref{NSGE}), and in the dispersion relation $\omega(k)$ of
the junction, which determines the voltage position $V_n$ of Fiske
resonances. To compare with the experimental data, we include the
frequency dependence of $\lambda_L (\omega) $ (material
dispersion), and relate the critical magnetic field $H_{c1}$ to
the effective size of a vortex in the quasi-local limit.

\subsection{Geometrical nonlocality}\label{geomnonlocalsec}

In a junction of finite length $L$ the translational invariance
along the junction is broken and, in general, the kernel
$Q(x,x^\prime)$ depends not only on $x-x^\prime$, but also on the
sum $x+x^\prime$ and the length $L$. However, in a finite
junction, deviations of the nonlocality kernel from the one of an
infinitely long junction are expected, mainly in the edge region
on the order of the magnetic screening length $\lambda_L$ (or
$\lambda_P =2 \lambda_L^ 2 /w$ in thin films). These deviations
can be neglected for the long junctions $L \gg \lambda_L$ ($L \gg
\lambda_P$) considered in the following. This assumption is
justified by the fairly good agreement obtained with the
experimental data presented in figures \ref{lengthAL},
\ref{FFS_AL}, \ref{length}, \ref{FFS_NB}, \ref{DISP_AL} and
\ref{DISP_NB}.

We distinguish between nonlocal effects due to stray fields (and
screening currents) inside and outside the junction electrodes.
These two contributions are referred as internal and geometrical
nonlocality, respectively.

(a) Let us first consider the theory for internal nonlocal effects
in bulk junctions of width $w\gg\lambda_L$, see.
Refs.~\onlinecite{Gur, AliSil}. In this case the  kernel in
Eq.~(\ref{NSGE}) is
\begin{equation}
Q(x)=(1/\pi \lambda_L)K_0(|x|/\lambda_L),
\end{equation}
where $K_0$ is a modified Bessel function, having a logarithmic
pole at $x\to0$ and exponentially decaying at large distances
\cite{AbrSteg}. Therefore, nonlocal effects become important here,
when $\lambda_L$ is large compared to the typical length scale
$l_\varphi$, on which the phase varies , i.e. in the case of
$\lambda_L\geq\lambda_J$. In the limit $\lambda_L \ll \lambda_J$
the local sine-Gordon equation is recovered.

Usually $\lambda_J$ is much larger than $\lambda_L$, because the
critical current density $j_c$ in the junction is much smaller
than the critical (depairing) current density
$j_d=2\Phi_0/(3\sqrt{3}\pi^2\mu_0\lambda_L^2\xi)$ in the bulk
superconductor ($\xi$ is the coherence length). Exceptions
are, for example, junctions in high-temperature superconductors
created by planar defects such as twins, stacking faults,
low-angle grain boundaries, etc \cite{GrainRev}. These structural
defects often form Josephson junctions with high $j_c$ and
therefore small $\lambda_J$. In terms of the critical current
density across the junction, the condition $\lambda_J < \lambda_L$
is fulfilled if $j_d/\kappa<j_c<j_d$, where $\kappa=\lambda_L/\xi$
is the Ginzburg-Landau parameter \cite{Gur}. Note that for extreme
type-II superconductors, $\kappa\gg1$, the relation
$\lambda_J<\lambda_L$ holds over a wide range of $j_c$.

Linearizing the wave equation (\ref{NSGE})
($\sin\varphi\simeq\varphi$), one obtains the dispersion relation
$\omega(k)$ for small-amplitude linear electromagnetic waves
$\varphi(x,t)=\varphi_0\exp(-{\rm i}\omega t+{\rm i} kx)$ (here
$|\varphi_0|\ll1$) propagating along a Josephson junction. From
the above kernel it follows that
\begin{equation}
\omega = \omega_p \left( 1 + \frac{k^2 \lambda_J^2}{\sqrt{1 + k^2
\lambda_L^2 }}   \right)^{\frac{1}{2}}  .
\end{equation}

(b) The geometrical nonlocality of an edge-type junction formed
between two thin superconducting films $w<\lambda_L$
(Fig.~\ref{allgeom}(a)) has been extensively studied
\cite{Kogan,IvanSob,MinSnap}. In this case, the stray fields
outside the junction area account for the entire electromagnetic
energy of the junction.  The typical length scale upon which
magnetic fields vary, is not the  London penetration depth
$\lambda_L$, but the Pearl penetration depth \cite{Pearl}
$\lambda_P=2 \lambda_L^2/w \gg \lambda_L$. The nonlocality kernel
is then given by \cite{Kogan}
\begin{eqnarray}\label{kernel1}
\nonumber Q(x)&=& \frac{4\lambda_L}{w}\int\frac{dkdq}{(2\pi)^2}
  \frac{\exp({\rm i} kx)}{\sqrt{k^2+q^2}[1+\lambda_P^2q^2]}
\\
&=&\frac{1}{2 \lambda_L} \left( {\rm H}_0 \left(
\frac{|x|}{\lambda_P} \right) - Y_0 \left(  \frac{|x|}{\lambda_P}
\right)   \right) ,
\end{eqnarray}
where ${\rm H}_0$ and $Y_0$ are the Struve and Bessel function of
the second kind\cite{AbrSteg}. As in the previous case, this
kernel has a logarithmic pole at $x\to0$, however at large
distances it decays as $\sim 1/x$. When the phase variation length
$l_\varphi$ is large compared to the magnetic screening length,
$\lambda_P\ll l_\varphi$, the characteristic vortex size is given
by ${\tilde \lambda}_J = l_\varphi = \lambda_J (\lambda_L/w)^{1/2}
\gg \lambda_J$. The limit of $\lambda_P \ll l_\varphi$ is then
equivalent to $\lambda_L \ll \lambda_J (w/\lambda_L)^{1/2}$. In
the opposite limit of $l_\varphi \ll \lambda_P$ or $\lambda_L \gg
\lambda_J (w / \lambda_L)^{1/2} $ the length scales for the tails
(${\tilde \lambda}_J $) and the core ($\lambda_J^2/ \lambda_L$) of
a Josephson vortex are different\cite{Kogan}.

(c) Ivanchenko's theory \cite{Ivan} applies to edge-type junctions
of width larger than the London penetration depth, $w \gtrsim
\lambda_L$. This model assumes that stray fields outside the film
affect only the surface, and that the interior of the junction is
described by the local theory. In this case the integral kernel
has the form \cite{Ivan}:
\begin{eqnarray}\label{kernel}
\nonumber Q(x)&=&\delta(x) +
\frac{4\lambda_L}{w}\int\frac{dkdq}{(2\pi)^2}
    \frac{\exp({\rm i} kx)}{\sqrt{k^2+q^2}[1+\lambda_L^2q^2]} =
  \\
&=& \delta (x) + \frac{1}{w} \left( {\bf H}_0 \left(
\frac{|x|}{\lambda_L} \right) - Y_0 \left(  \frac{|x|}{\lambda_L}
\right)   \right)
\end{eqnarray}
Note that the second term is identical to the kernel of
Eq.~(\ref{kernel1}), corresponding to the case of a thin film
junction, but the Pearl penetration depth $\lambda_P$ is replaced
by the London penetration depth $\lambda_L$. As in the previous
case, when $l_\varphi$ is large compared to the typical length
scale of the kernel, $\lambda_L\ll l_\varphi$,
 the vortex size is given by $\tilde \lambda_J=l_\varphi =\lambda_J
(1+\lambda_L/w)^{1/2}$.

In the limit of $l_\varphi \gg \lambda_L$ for cases (b) and (c),
Eq.~(\ref{NSGE}) does not transform exactly into the local
sine-Gordon equation with effective length scale ${\tilde
\lambda}_J$, because $Q(x)$ does not correspond to $\delta(x)$ for
$\lambda_{P/L} \rightarrow 0$. Nevertheless, the locality of the
kernel suggests that in this limit an effective local sine-Gordon
equation with renormalized Josephson penetration depth ${\tilde
\lambda}_J$ should be qualitatively correct. Then one could use
the expressions for $H_{c1}$ and $\bar{c}= \omega_{\rm p}
\lambda_J$ from local theory, if $\lambda_J$ is replaced by a new
length $\tilde\lambda_J$, see below.

In the absence of dissipation and bias current, using the kernel
(\ref{kernel}) we obtain the dispersion relation for this case as
\begin{equation}\label{disp1}
    \omega(k)=\omega_p\sqrt{1+k^2\lambda_J^2+ \frac{4k^2\lambda_J^2 \Lambda}{\pi
    w\sqrt{k^2\Lambda^2-4}}
    \arccos\left(\frac{2}{|k|\Lambda}\right)}
\end{equation}
for $k\Lambda\geq2$ and
\begin{equation}\label{disp2}
    \omega(k)=\omega_p\sqrt{1+k^2\lambda_J^2+ \frac{4k^2\lambda_J^2 \Lambda}{\pi
    w\sqrt{4-k^2\Lambda^2}}\
    \arccosh\left(\frac{2}{|k|\Lambda}\right)}
\end{equation}
for $k\Lambda<2$.

(d) The theory developed by Lomtev and Kuzovlev \cite{LomKuz} is
the most general as it contains the special cases (a)-(c)
discussed above. It assumes that the phase $\varphi(x)$ does not
vary over the width $w$. The kernel is
\begin{equation}
Q (x) = \frac{1}{\pi \lambda_L} K_0 \left( \frac{|x|}{\lambda_L}
\right) + \frac{2}{\pi w \lambda_L^3} \int_0^\infty \frac{dq J_0
(qx) }{ \kappa^3 ( \kappa  + q {\rm coth}(\kappa w/2) ) },
\end{equation}
where $\kappa = (q^2 + 1/\lambda_L^2)^{1/2} $, $J_0$ is the Bessel
function of the first kind. In this case the dispersion relation
in the absence of perturbation terms in Eq.~(\ref{NSGE}) is
\cite{LomKuz}
\begin{equation}\label{dispLK}
    \omega(k)=\omega_p\sqrt{1+\frac{k^2 \lambda_J^2}{\sqrt{1 + k^2 \lambda_L^2}}
    + \frac{k^2\lambda_J^2}{\pi\lambda_L}F(k)} \ \ ,
\end{equation}
where
\begin{equation}\label{dispLK1}
    F(k)=\frac{2}{w\lambda_L^2}\int\limits_k^\infty dq\frac{1}{\kappa^3}
    \frac{1}{\kappa+k\coth(\kappa w/2)} \frac{1}{(q^2-k^2)^{1/2}}\
    \ .
\end{equation}

This theory has also been generalized to describe ramp-type
junctions \cite{Lomtev} (Fig.~\ref{allgeom}(b)).

Let us quote typical parameters for our low-$T_c$ long Josephson
junctions. For the Nb-AlO$_x$-Nb junctions considered in this
paper $\lambda_J\simeq25\ \mu$m, and $\lambda_L\simeq90\ $nm at
$T=4.2$ K, and for Al-AlO$_x$-Al junctions $\lambda_J\simeq56\
\mu$m and $\lambda_L\simeq15\ $nm at $T$=0.3 K, and
$\lambda_J\simeq68\ \mu$m and $\lambda_L\simeq17\ $nm at $T$=0.9
K. For these parameters, the condition $\lambda_L\ll\lambda_J$ is
fulfilled in all cases, i.e. the internal nonlocal effects as
discussed in case (a) are negligibly small.  For Al-AlO$_x$-Al
junctions $w\approx 0.1-0.8\ \mu$m, and for Nb-AlO$_x$-Nb
junctions $w\approx 0.3-4.2\ \mu$m, which is significantly larger
than $\lambda_L$. It is reasonable to expect that our junctions
can be approximately described by the theory presented in
Ref.~\onlinecite{Ivan}. In the following analysis, we compare our
experimental data with Ivanchenko's theory \cite{Ivan}, case (c)
and with the more general theory of Lomtev and Kuzovlev
\cite{LomKuz}, case (d).

\subsection{Material dispersion}

We will see in Sec. \ref{Fisketheory}, that for small frequencies
the nonlocal dispersion relation predicts a strong decrease of the
spacing between the Fiske steps $\Delta V_{FS\ n}=V_{n} - V_{n-1}$
with increasing $n$ (cf. Eqs.~(\ref{disp1}), (\ref{disp2}),
(\ref{dispLK}) and (\ref{dispFS})). For large wave numbers, this
effect is of the same order of magnitude as the reduction of the
spacing between Fiske steps due to material dispersion \cite{Lee}.
Material dispersion arises from the frequency dependence of the
complex conductivity and surface impedance of the superconducting
electrodes \cite{Mattis, Tinkham}. It leads to a frequency
dependence of $\lambda_L$ which becomes significant at frequencies
on the order of the superconducting energy gap frequency $f_{\rm
gap}=\Delta/h$ ($\Delta$ is the superconducting energy gap). The
gap frequency of the Nb-AlO$_x$-Nb junctions is $f_{\rm
gap}\simeq650\ $GHz (at $4.2\ $K) and that of Al-AlO$_x$-Al is
$f_{\rm gap}\simeq85\ $GHz (at $0.3\ $K). The frequency dependence
of $\lambda_L$ can be written as\cite{Mattis,Lee}
\begin{equation}\label{Mattisformula}
\left(\frac{\lambda_L(0)}{\lambda_L(\omega)}\right)^2 =
\sqrt{\pi\Delta}
\int\limits_{\Delta-\hbar\omega}^{\Delta}dx\left[1-2
f\left(x+\hbar\omega\right)\right] \frac{x^2 + \Delta^2 +
x\hbar\omega} {\sqrt{\Delta^2 - x^2}\sqrt{(x +\hbar\omega)^2 -
\Delta^2}},
\end{equation}
where $f(x)=1/(1+e^{x/k_B T})$ is the Fermi-Dirac distribution
function, $k_B$ is Boltzmann's constant.

\subsection{Fiske steps}\label{Fisketheory}

Let us consider a long junction of finite length $L$ in the
direction of wave propagation. Because of reflection at the edges,
it behaves like a resonant transmission line and supports cavity
modes of the electromagnetic field. The theory of such resonances
(Fiske resonances) in the local case was developed by Kulik
\cite{KulikFiske,Kulik}. The nonlocal theory for semi-infinite,
and finite Josephson junctions \cite{Kogansemi,Privat} suggest
that for case (c) of Ivanchenko \cite{Ivan}, in the presence of an
external magnetic field $H_{e}$, we can use the boundary condition
\begin{equation}\label{boundCond}
\varphi_x(0)=\varphi_x(L)=\frac{2\pi\Lambda}{\Phi_0}H_{e}.
\end{equation}
Then, the Josephson phase is written as
\begin{equation}
    \varphi=\omega t - k x + \varphi_1(x,t)
\end{equation}
where $\omega=2\pi V/\Phi_0$ and $k= 4\pi\lambda_LH_{e}/\Phi_0$.
As a first approximation, we consider $\varphi_1(x,t)$ as a small
perturbation. By inserting this into Eq.~(\ref{NSGE}) we obtain
\begin{equation}\label{NSGEFiske}
\lambda_J^2\frac{\partial} {\partial x}\int
dx'Q(x-x')\varphi_{1,x'}(x')-
\omega_p^{-2}\varphi_{1,tt}-\omega_p^{-1}\alpha\varphi_{1,t}=
\sin(\omega t- kx),
\end{equation}
where we neglect $\varphi_1(x,t)$ in the sine term, and use the
assumption concerning the nonlocality kernel made at the beginning
of Sec.~\ref{geomnonlocalsec}. Furthermore, $\varphi_1(x,t)$ can
be expanded in terms of the normal modes of the junction
\begin{equation}\label{varphi1}
    \varphi_1(x,t)={\rm Im}\left\{\sum\limits_{n=0}^\infty g_n
    e^{\rm i \omega_n t} \cos k_nx \right\},
\end{equation}
where $g_n$ are complex numbers, $k_n=n\pi/L$. This choice for the
$x$ dependence implies the boundary conditions
$\varphi_{1,x}(0)=\varphi_{1,x}(L)=0$, which is consistent with
(\ref{boundCond}). The frequency $\omega_n$ is determined from the
dispersion relation of Eq.~(\ref{NSGEFiske}), namely
$\sqrt{\omega^2(k)-\omega_p^2}$, $\omega(k)$ is determined in
Sec.~\ref{geomnonlocalsec}. By inserting the expression
(\ref{varphi1}) for $\varphi_1$ into Eq.~(\ref{NSGEFiske}), and
solving the set of the equations for $g_n$, we obtain the
following expression for the position of the Fiske resonances
\begin{equation}\label{dispFS}
V_n=\frac{\hbar}{2e}\sqrt{\omega^2\left(k_n\right)-\omega_p^2}.
\end{equation}
In the local case ($Q(x)=\delta(x)$, $\omega^2(k)=\omega_p^2 +
\bar{c} k^2$) we obtain Eq.~(\ref{Fiskelocal}).

Using Eqs.~(\ref{disp1}), (\ref{disp2}),
(\ref{dispLK})-(\ref{Mattisformula}) and (\ref{dispFS}), in
Fig.~\ref{FFS_AL} we plot the dependence of the first Fiske step
voltage $V_1$ on the junction width $w$ for Al-AlO$_x$-Al
junctions, and in Fig.~\ref{FFS_NB} that for Nb-AlO$_x$-Nb
junctions. The only fitting parameter for all curves is the
Swihart velocity. For the Al-AlO$_x$-Al junctions we found $\bar
c\simeq8\times 10^6$ m/s and for Nb-AlO$_x$-Nb junctions $\bar
c\simeq7\times 10^6$ m/s. The Swihart velocity found using the
theory of case \cite{Ivan} (c) and that of case\cite{LomKuz} (d)
differ by about 1\%. This small difference indicates that internal
nonlocal effects, taken into account in case (d), are negligibly
small in our junctions.

\begin{figure}
  \includegraphics[scale=1.26]{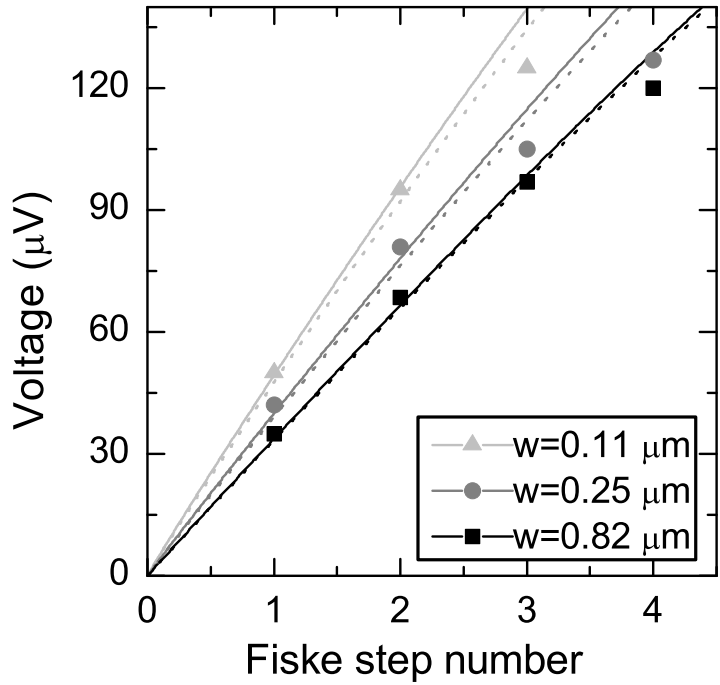}
  \caption{Comparison of experimentally obtained Fiske step
  voltages for Al-AlO$_x$-Al junctions of different widths (symbols)
  with the dispersion curve calculated from nonlocal theories
  taking into account material dispersion (lines).
  Triangles correspond to the junction of width $w=0.11\,\mu$m,
  circles to $w=0.25\,\mu$m and rectangles to $w=0.82\,\mu$m. The solid
  line corresponds to the theory of case \cite{Ivan} (c) and the dashed line
  to case \cite{LomKuz} (d).}
  \label{DISP_AL}
\end{figure}

\begin{figure}
  \includegraphics[scale=1.26]{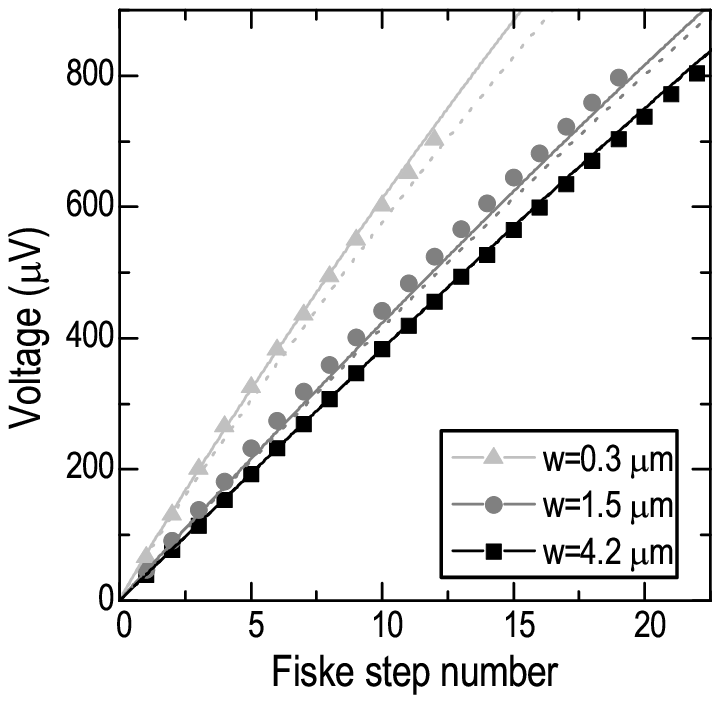}
  \caption{Comparison of experimentally obtained Fiske step
  voltages for Nb-AlO$_x$-Nb junctions of different widths (symbols)
  with the dispersion curve calculated from nonlocal theories
  taking into account material dispersion (lines).
  Triangles correspond to the junction of width $w=0.3\,\mu$m,
  circles to $w=1.5\,\mu$m and rectangles to $w=4.2\,\mu$m. The solid
  line corresponds to the theory of case \cite{Ivan} (c) and the dashed line
  to case \cite{LomKuz} (d).}
  \label{DISP_NB}
\end{figure}

In Figs.~\ref{DISP_AL} and \ref{DISP_NB} we compare the measured
(symbols) voltage positions $V_n$ of the Fiske steps to
theoretical curves calculated from Eqs.~(\ref{disp1}),
(\ref{disp2}), (\ref{dispLK})-(\ref{Mattisformula}) and
(\ref{dispFS}), where both nonlocality and material dispersion
have been taken into account. Solid lines correspond to the theory
of case\cite{Ivan} (c), and dashed lines to the theory of
case\cite{LomKuz} (d). The data for Al junctions is presented in
Fig.~\ref{DISP_AL}, and for Nb junctions presented in
Fig.~\ref{DISP_NB}. The difference between the curves according to
the two theories decreases with increasing the width of the
junctions. For the Nb-AlO$_x$-Nb junction of width $4.2\ \mu$m the
difference cannot be distinguished.

Thus, our experimental data for Fiske steps in both Al-AlO$_x$-Al
and Nb-AlO$_x$-Nb junctions are in good agreement with theory
\cite{Ivan,LomKuz}.

\subsection{\label{vortexsize}Critical magnetic fields and vortex size}

For a junction placed in an external magnetic field the
integro-differential equation~(\ref{NSGE}) contains an additional
Meissner current $j_m (x,y,z)$, which is also induced in the
absence of the junction and depends on the size and the shape of
the electrodes (demagnetization effects). If the dynamics of the
phase inside a small surface region of size $\sim \lambda_L \ll L,
{\tilde \lambda}_J$ is disregarded, the geometry dependent current
$j_m$ can formally be taken into account by supplying the nonlocal
wave equation for the case of infinite length with appropriate
(nonlocal) boundary conditions. Physically, these boundary
conditions reflect edge capacitances and inductances, and take
into account the stray field in the $xy$-plane for $-w/2\le z \le
w/2$, in addition to the stray fields above and below the
superconducting leads, which are contained in the nonlocal kernel
$Q(x)$. These boundary conditions depend on the global geometry of
the sample and the magnetic screening length in addition to the
kernel $Q$. The magnetic penetration field $H_{c1}$ is therefore
more difficult to interpret theoretically than the Fiske steps,
which follow from the kernel $Q$ alone.

Although in general the solution is difficult to find, in the
physically relevant case (c) for $\lambda_L \ll l_\varphi$, where
the Eq.~(\ref{NSGE}) transforms into the local sine-Gordon
equation with an effective length scale ${\tilde \lambda}_J$, the
expressions for $H_{c1}$ and $\bar{c}= \omega_p \lambda_J$ from
the local theory can be used. In doing so the correct magnetic
screening length, e.g. $\lambda_L$ for $w \gtrsim \lambda_L$, and
the effective vortex size ${\tilde \lambda}_J = \lambda_J ( 1 +
\lambda_L /w)^{1/2}$ from the effective local theory has to be
used. Assuming that bulk material properties like $j_c$ do not
depend on the junction width $w$, it is possible to obtain the
scaling relation\cite{Koval}
\begin{equation}
\label{scaling} \bar{c} (w) \sim \frac{1}{H_{c1}^{1/3} (w)}
\end{equation}
between these quantities, when the width $w$ is changed, which is
well satisfied experimentally \cite{Koval}. Even if the effective local
theory provides a good approximation for the static case,
 it might nevertheless fail to describe in the dynamic case some significant
properties of the junction, such as the possibility of Cherenkov
radiation above a critical velocity (similarly for the internal
nonlocality \cite{MinSnapcheren}).

We will now verify the correct choice of the length scale ${\tilde
\lambda}_J$ of a Josephson vortex using a variational approach. As
the variation parameter we take the vortex size ${\tilde
\lambda}_J$, and minimize the Lagrange function of the unperturbed
($\alpha=\gamma=0$) nonlocal sine-Gordon equation (\ref{NSGE})
\begin{eqnarray}\label{lagrange}
\frac{{\cal L}}{  E_J w} &=&   \int dx  \left(\frac{1}{2}
\frac{1}{\omega_p^2} {\varphi}_t^2  -  (1 - \cos \varphi) \right)-
\frac{1}{2} \lambda_{J}^2 \int dx dy~  \varphi_x (x)  \varphi_y
(y) Q (x-y) \\
\end{eqnarray}
As trial functions we took the two limiting solutions of
Eq.~(\ref{NSGE}). For $w\to\infty$, the vortex solution of
Eq.~(\ref{NSGE}) is an exponential kink \cite{lamb}
\begin{equation}\label{expkink}
\varphi(x,\tilde\lambda_J)=4\,{\rm
arctan}\left(\exp\frac{x}{\tilde\lambda_J}\right),
\end{equation}
The other limit is $w\to0$, for which the one vortex solution
decays algebraically \cite{Kogan}:
\begin{equation}\label{lorkink}
\varphi(x,\tilde\lambda_J)=2\,{\rm arctan}\frac{x}{
\tilde\lambda_J}+\pi.
\end{equation}
Here $\tilde\lambda_J$ is considered as a free variational
parameter corresponding to the characteristic size of the vortex.
The results for Al-AlO$_x$-Al junctions calculated from the first
critical field $H_{c1}$ at $T=0.9$ K are presented in
Fig.~\ref{lengthAL} by symbols. The lines correspond to
variational results for the theories of Refs.~\onlinecite{Ivan}
and \onlinecite{LomKuz} with the above two ansatzes. Each curve
was fitted to the experimental data for junction widths in the
range $0.7-0.8~\mu$m. The prefactors for different theories differ
by 3 \%. Figure~\ref{length} presents the characteristic vortex
size $\tilde \lambda_J$ as a function of junction width for
Nb-AlO$_x$-Nb junctions, for $T=4.2$ K and $\lambda_L=90\ $nm. In
this figure the curves are fitted to the experimental data point
at width $w=4.2\mu$m. One notes that in both cases all four curves
overlap with each other.

The agreement between experimental data (symbols) and theoretical
estimations (lines) in Figs.~\ref{length}, \ref{lengthAL} is
rather poor. We suppose this is due to the fact that none of the
theories properly consider the boundary conditions at the edges of
a junction. Deriving appropriate boundary conditions for realistic
junction geometries in the nonlocal case remains an unsolved
problem, which requires further work.

\subsection{Vortex mass}

Finally, we would like to discuss the effect of nonlocal
electrodynamics on the dynamical mass and quantum behaviour of
Josephson vortices. We note that the first experiment with a
vortex in the quantum regime was presented in
Ref.~\onlinecite{ustinovvortex}. The (nonrelativistic) mass $m_F$
of a Josephson vortex, which appears in the equation of motion
\begin{equation}
m_F {\ddot q} + \frac{\partial {\cal H}_{\rm ext}}{\partial q} = 0
\end{equation}
for the vortex center of mass coordinate $q$ in an external
potential ${\cal H}_{\rm ext}$, is obtained by expanding the
Lagrange function
\begin{equation}
{\cal L} \approx \frac{1}{2} m_F v^2- H_{\rm ext}+{\rm const}
\end{equation}
of Eq.~(\ref{lagrange}) in terms of $v \ll \bar{c}$. The spacing
$\hbar \omega_0$ between the energy levels (and, thus, zero point
fluctuations) is then given by
\begin{equation}
\omega_0^2  = \frac{1}{2 m_F} \frac{\partial^2 H_{\rm ext}}{
\partial q^2}.
\end{equation}
The condition $\hbar \omega_0 > k_B T$ is required for dominantly
quantum behaviour of the fluxon. In the limit $m_F \rightarrow 0$,
the spectrum becomes discrete, provided that the energy scale of
the potential is fixed.

The external potential ${\cal H}_{\rm ext}$ for a vortex in a
Josephson junctions can be, for example, induced by an external
magnetic field or by presence of a micro-short (micro-resistor).
In the first case, the external magnetic field creates a screening
current flowing through the junction. Provided that the
nonlocality of the magnetic screening at the edges can be
neglected or, equivalently, Eq.~(\ref{boundCond}) for the boundary
condition is justified, the current spreads homogenously over the
width of the junction. Hence, the external potential ${\cal
H}_{\rm ext}$ is proportional to the junction width
\cite{MonMart}. In the second case, the inhomogeneity leads to a
local change in the Josephson energy. If the magnitude of the
inhomogeneity is constant across the junction, the total change in
the Josephson energy is proportional to $w$. As we see, in both
situations the external potential ${\cal H}_{\rm ext}$ scales with
$w$, hence, for characterizing the degree of quantum behavior, the
use of the {\em specific} mass ${\tilde m}_F = m_F/w$ for a vortex
is more appropriate than using the mass $m_F$.

For local theory ($\lambda_J \gg w \gg \lambda_L$), the mass of a
vortex of form (\ref{expkink}) is given by
\begin{equation}
m_F = \frac{8 E_J w}{\omega_p^2\lambda_J},
\end{equation}
which vanishes in the limit $w \rightarrow 0$, whereas the
specific mass ${\tilde m}_F = m_F /w$ is constant.

Similarly,  in the nonlocal case  we make use of the ansatz
$\phi(x,t,v) = {\tilde \phi} ( \tilde x,v)$, where $\tilde x =
(x-vt)/\tilde\lambda_J(v)$ and ${\tilde \lambda}_J (v)$ is the
characteristic length scale of the vortex, which in general
depends on the velocity (cf. the Lorentz-factor
$1/\sqrt{1-(v/\bar{c})^2}$ in the local case), and which can be
determined from the variational ansatz. As the Lorentz invariance
is broken in the nonlocal case, we do not know a priori the
functional dependence of $\phi$ on $v$, and an additional explicit
dependence on $v$ is possible, if the shape of the vortex depends
on $v$. From (\ref{lagrange}) we obtain
\begin{eqnarray}
\frac{ {\cal L}}{  E_J w }  &=& \frac{1}{2\omega_p^2} \frac{1}{
\tilde\lambda_J} I_1 v^2 + \tilde\lambda_J I_2 +
\frac{\lambda_{J}^2}{2} } ~{ I_3 \left( {\tilde \lambda}_J (v),
\lambda_L, w \right),
\end{eqnarray}
where
\begin{eqnarray}
I_1 &=&  \int d {\tilde x}\ ( {\tilde  \varphi}^\prime ( {\tilde
x} ) )^2\label{integralI1};
\\
I_2 &=&- \int d {\tilde x}\ ( 1 - \cos {\tilde \varphi}  );
\\
I_3 &=&- \int d {\tilde x}\ d {\tilde y}\ {\tilde \varphi}'
({\tilde x})  {\tilde \varphi}' ({\tilde y}) Q( {\tilde \lambda}_J
|{\tilde x} - {\tilde y}|).
\end{eqnarray}
The effective mass in the nonlocal case is obtained by expanding
${\cal L} (v)$ into second order. Neglecting the explicit
dependence on $v$, and expanding
\begin{equation}
{\tilde \lambda}_J (v) = {\tilde \lambda}_{J0} + \frac{1}{2}
\frac{d^2\tilde\lambda_J}{dv^2} {v}^2 = {\tilde \lambda}_{J0} +
\frac{1}{2} \frac{v^2}{{\tilde c}^2},
\end{equation}
one obtains
\begin{equation}
m_f = \frac{E_J w}{\omega^2_p}\frac{I_1}{\tilde \lambda_{J0}} +
\frac{E_J w}{2} \frac{d^2\tilde\lambda_J}{dv^2} \left(2 I_2 +
\lambda_{J}^2 \frac{\partial
  I_3 ({\tilde \lambda}_J={\tilde \lambda}_{J0})}{\partial {\tilde
  \lambda}_J}\right).
\label{general}
\end{equation}
Here we again apply the variational approach as we did to obtain
the vortex size. Then the expression in brackets is the Lagrange
equation for the variation parameter $\tilde\lambda_{J0}$, and is
therefore equal to zero. Thus the effective mass for the trial
function of the form (\ref{expkink}) can explicitly be written as
\begin{equation}
m_f=\frac{8E_J w}{\omega^2_p \tilde \lambda_{J0}}.
\label{specificmass}
\end{equation}

Since $\tilde\lambda_J$ increases with $w$, both the effective
mass $m_f$ and the specific mass $m_f/w$ (responsible for the
energy scaling, if ${\cal H}_{\rm ext} \sim w$) decrease with
decreasing the junction width. From Eq.~(\ref{specificmass}) one
notes that the inverse of the specific mass scales with the
characteristic length $\tilde \lambda_{J0}$ (see
Figs.~\ref{lengthAL} and \ref{length}).

Let us emphasize again that the possibility of a vortex behaving
as a macroscopic quantum particle, when reducing the width $w$ of
the junction, not only depends on the scaling of the (kinetic)
mass $m_F$, but more exactly the ratio of the kinetic energy to the external
potential ${\cal H}_{\rm ext}$, which has to be determined for
each specific experimental situation.

\section{Conclusion}\label{conclusion}

The current-voltage characteristics and critical current versus
field patterns of narrow long Josephson junctions show a strong
dependence on the junction width $w$. This cannot be explained
using the conventional theory based on the local sine-Gordon
model. Our experimental data are well described by the nonlocal
theory originally developed by Ivanchenko \cite{Ivan}, and later
extended by Lomtev and Kuzovlev \cite{LomKuz}. According to these
models, the electrodynamic and static properties of a junction
depend on the junction width. Our experimental data for Fiske
steps observed in both Al-AlO$_x$-Al and Nb-AlO$_x$-Nb long narrow
junctions are in very good quantitative agreement with these
theories. The dependence of the first critical field on the width
of the junction allows qualitative estimates of the characteristic
vortex size, which is influenced by nonlocal effects. A proper
description of the static effects requires a better treatment of
the boundary conditions at the edges of a junction.

We also calculate the specific mass of a vortex in the nonlocal
case. According to local theory, the specific mass does not depend
on the junction width $w$. By contrast, the nonlocal theory predicts
that the specific mass of a vortex decreases with decreasing
junction width.

\section{Acknowledgements}
We would like to thank M.\ V.\ Fistul for useful discussions and
A. Price for careful reading of this manuscript.


\begin{thebibliography}{38}
\expandafter\ifx\csname
natexlab\endcsname\relax\def\natexlab#1{#1}\fi
\expandafter\ifx\csname bibnamefont\endcsname\relax
  \def\bibnamefont#1{#1}\fi
\expandafter\ifx\csname bibfnamefont\endcsname\relax
  \def\bibfnamefont#1{#1}\fi
\expandafter\ifx\csname citenamefont\endcsname\relax
  \def\citenamefont#1{#1}\fi
\expandafter\ifx\csname url\endcsname\relax
  \def\url#1{\texttt{#1}}\fi
\expandafter\ifx\csname
urlprefix\endcsname\relax\def\urlprefix{URL }\fi
\providecommand{\bibinfo}[2]{#2}
\providecommand{\eprint}[2][]{\url{#2}}

\bibitem[{\citenamefont{Kulik and Yanson}(1972)}]{Yanson}
\bibinfo{author}{\bibfnamefont{I.~O.} \bibnamefont{Kulik}} \bibnamefont{and}
  \bibinfo{author}{\bibfnamefont{I.~K.} \bibnamefont{Yanson}},
  \emph{\bibinfo{title}{The Josephson Effect in Superconducting Tunnel
  Structures}} (\bibinfo{publisher}{Keter Press}, \bibinfo{address}{Jerusalem},
  \bibinfo{year}{1972}).

\bibitem[{\citenamefont{Barone and Patern\'{o}}(1982)}]{BarPat}
\bibinfo{author}{\bibfnamefont{A.}~\bibnamefont{Barone}} \bibnamefont{and}
  \bibinfo{author}{\bibfnamefont{G.}~\bibnamefont{Patern\'{o}}},
  \emph{\bibinfo{title}{Physics and Applications of the Josephson Effect}}
  (\bibinfo{publisher}{Willey}, \bibinfo{address}{New York},
  \bibinfo{year}{1982}).

\bibitem[{\citenamefont{Likharev}(1986)}]{Likharev}
\bibinfo{author}{\bibfnamefont{K.~K.} \bibnamefont{Likharev}},
  \emph{\bibinfo{title}{Dynamics of Josephson Junctions and Circuits}}
  (\bibinfo{publisher}{Gordon and Breach}, \bibinfo{address}{New York},
  \bibinfo{year}{1986}).

\bibitem[{\citenamefont{Ustinov}(1998)}]{Ustinov}
\bibinfo{author}{\bibfnamefont{A.~V.} \bibnamefont{Ustinov}},
  \bibinfo{journal}{Physica\ D} \textbf{\bibinfo{volume}{123}},
  \bibinfo{pages}{315} (\bibinfo{year}{1998}).

\bibitem[{\citenamefont{A.Wallraff et~al.}(2003)\citenamefont{A.Wallraff,
  A.Lukashenko, J.Lisenfeld, A.Kemp, M.V.Fistul, Koval, and
  Ustinov}}]{ustinovvortex}
\bibinfo{author}{\bibnamefont{A.Wallraff}},
  \bibinfo{author}{\bibnamefont{A.Lukashenko}},
  \bibinfo{author}{\bibnamefont{J.Lisenfeld}},
  \bibinfo{author}{\bibnamefont{A.Kemp}},
  \bibinfo{author}{\bibnamefont{M.V.Fistul}},
  \bibinfo{author}{\bibfnamefont{Y.}~\bibnamefont{Koval}}, \bibnamefont{and}
  \bibinfo{author}{\bibfnamefont{A.}~\bibnamefont{Ustinov}},
  \bibinfo{journal}{Nature} \textbf{\bibinfo{volume}{425}},
  \bibinfo{pages}{155} (\bibinfo{year}{2003}).

\bibitem[{\citenamefont{Fistul et~al.}(2003)\citenamefont{Fistul, Wallraff,
  Koval, Lukashenko, Malomed, and Ustinov}}]{Fistul}
\bibinfo{author}{\bibfnamefont{M.~V.} \bibnamefont{Fistul}},
  \bibinfo{author}{\bibfnamefont{A.}~\bibnamefont{Wallraff}},
  \bibinfo{author}{\bibfnamefont{Y.}~\bibnamefont{Koval}},
  \bibinfo{author}{\bibfnamefont{A.}~\bibnamefont{Lukashenko}},
  \bibinfo{author}{\bibfnamefont{B.~A.} \bibnamefont{Malomed}},
  \bibnamefont{and} \bibinfo{author}{\bibfnamefont{A.~V.}
  \bibnamefont{Ustinov}}, \bibinfo{journal}{Phys. Rev. Lett.}
  \textbf{\bibinfo{volume}{91}}, \bibinfo{pages}{257004}
  (\bibinfo{year}{2003}).

\bibitem[{\citenamefont{Hermon et~al.}(1994)\citenamefont{Hermon, Stern, and
  Ben-Jacob}}]{Hermon}
\bibinfo{author}{\bibfnamefont{Z.}~\bibnamefont{Hermon}},
  \bibinfo{author}{\bibfnamefont{A.}~\bibnamefont{Stern}}, \bibnamefont{and}
  \bibinfo{author}{\bibfnamefont{E.}~\bibnamefont{Ben-Jacob}},
  \bibinfo{journal}{Phys.\ Rev.\ B} \textbf{\bibinfo{volume}{49}},
  \bibinfo{pages}{9757} (\bibinfo{year}{1994}).

\bibitem[{\citenamefont{Kato and Imada}(1996)}]{Kato}
\bibinfo{author}{\bibfnamefont{T.}~\bibnamefont{Kato}} \bibnamefont{and}
  \bibinfo{author}{\bibfnamefont{M.}~\bibnamefont{Imada}},
  \bibinfo{journal}{J.\ Phys.\ Soc.\ Jpn.} \textbf{\bibinfo{volume}{65}},
  \bibinfo{pages}{2963} (\bibinfo{year}{1996}).

\bibitem[{\citenamefont{Shnirman et~al.}(1997)\citenamefont{Shnirman,
  Ben-Jacob, and Malomed}}]{Shnirman}
\bibinfo{author}{\bibfnamefont{A.}~\bibnamefont{Shnirman}},
  \bibinfo{author}{\bibfnamefont{E.}~\bibnamefont{Ben-Jacob}},
  \bibnamefont{and} \bibinfo{author}{\bibfnamefont{B.}
  \bibnamefont{Malomed}}, \bibinfo{journal}{Phys.\ Rev.\ B}
  \textbf{\bibinfo{volume}{56}}, \bibinfo{pages}{14677} (\bibinfo{year}{1997}).

\bibitem[{\citenamefont{Koval et~al.}(1999)\citenamefont{Koval, Wallraff,
  Fistul, Thyssen, Kohlstedt, and Ustinov}}]{Koval}
\bibinfo{author}{\bibfnamefont{Y.}~\bibnamefont{Koval}},
  \bibinfo{author}{\bibfnamefont{A.}~\bibnamefont{Wallraff}},
  \bibinfo{author}{\bibfnamefont{M.}~\bibnamefont{Fistul}},
  \bibinfo{author}{\bibfnamefont{N.}~\bibnamefont{Thyssen}},
  \bibinfo{author}{\bibfnamefont{H.}~\bibnamefont{Kohlstedt}},
  \bibnamefont{and} \bibinfo{author}{\bibfnamefont{A.~V.}
  \bibnamefont{Ustinov}}, \bibinfo{journal}{IEEE Trans. Appl. Supercond.}
  \textbf{\bibinfo{volume}{9}}, \bibinfo{pages}{3957} (\bibinfo{year}{1999}).

\bibitem[{\citenamefont{Mattis and Bardeen}(1958)}]{Mattis}
\bibinfo{author}{\bibfnamefont{D.~C.} \bibnamefont{Mattis}} \bibnamefont{and}
  \bibinfo{author}{\bibfnamefont{J.}~\bibnamefont{Bardeen}},
  \bibinfo{journal}{Phys.\ Rev.} \textbf{\bibinfo{volume}{111}},
  \bibinfo{pages}{412} (\bibinfo{year}{1958}).

\bibitem[{\citenamefont{Lee and Brafknecht}(1992)}]{Lee}
\bibinfo{author}{\bibfnamefont{G.~S.} \bibnamefont{Lee}} \bibnamefont{and}
  \bibinfo{author}{\bibfnamefont{A.~T.} \bibnamefont{Brafknecht}},
  \bibinfo{journal}{IEEE\ Trans.\ Appl.\ Superconduct.}
  \textbf{\bibinfo{volume}{2}}, \bibinfo{pages}{67} (\bibinfo{year}{1992}).

\bibitem[{\citenamefont{Tinkham}(1996)}]{Tinkham}
\bibinfo{author}{\bibfnamefont{M.}~\bibnamefont{Tinkham}},
  \emph{\bibinfo{title}{Introduction to Superconductivity}}
  (\bibinfo{publisher}{McGraw-Hill}, \bibinfo{address}{New York},
  \bibinfo{year}{1996}), \bibinfo{edition}{2nd} ed.

\bibitem[{\citenamefont{Gurevich}(1992)}]{Gur}
\bibinfo{author}{\bibfnamefont{A.}~\bibnamefont{Gurevich}},
  \bibinfo{journal}{Phys.\ Rev.\ B} \textbf{\bibinfo{volume}{46}},
  \bibinfo{pages}{R3187} (\bibinfo{year}{1992}).

\bibitem[{\citenamefont{Aliev et~al.}(1992)\citenamefont{Aliev, Silin, and
  Uryupin}}]{AliSil}
\bibinfo{author}{\bibfnamefont{Y.~M.} \bibnamefont{Aliev}},
  \bibinfo{author}{\bibfnamefont{V.~P.} \bibnamefont{Silin}}, \bibnamefont{and}
  \bibinfo{author}{\bibfnamefont{S.~A.} \bibnamefont{Uryupin}},
  \bibinfo{journal}{Superconductivity} \textbf{\bibinfo{volume}{5}},
  \bibinfo{pages}{1992} (\bibinfo{year}{1992}).

\bibitem[{\citenamefont{Lapir et~al.}(1975)\citenamefont{Lapir, Likharev,
  Maslova, and Semenov}}]{Lapir}
\bibinfo{author}{\bibfnamefont{G.~M.} \bibnamefont{Lapir}},
  \bibinfo{author}{\bibfnamefont{K.~K.} \bibnamefont{Likharev}},
  \bibinfo{author}{\bibfnamefont{L.~A.} \bibnamefont{Maslova}},
  \bibnamefont{and} \bibinfo{author}{\bibfnamefont{V.~K.}
  \bibnamefont{Semenov}}, \bibinfo{journal}{Fiz.\ Niz.\ Temp.\ (Sov.\ J.\ Low\
  Temp.\ Phys.)} \textbf{\bibinfo{volume}{1}}, \bibinfo{pages}{1235}
  (\bibinfo{year}{1975}).

\bibitem[{\citenamefont{Kupriyanov et~al.}(1976)\citenamefont{Kupriyanov,
  Likharev, and Semenov}}]{Kupriyanov}
\bibinfo{author}{\bibfnamefont{M.~Y.} \bibnamefont{Kupriyanov}},
  \bibinfo{author}{\bibfnamefont{K.~K.} \bibnamefont{Likharev}},
  \bibnamefont{and} \bibinfo{author}{\bibfnamefont{V.~K.}
  \bibnamefont{Semenov}}, \bibinfo{journal}{Fiz.\ Niz.\ Temp.\ (Sov.\ J.\ Low\
  Temp.\ Phys.)} \textbf{\bibinfo{volume}{2}}, \bibinfo{pages}{706}
  (\bibinfo{year}{1976}).

\bibitem[{\citenamefont{Ivanchenko and Soboleva}(1990)}]{IvanSob}
\bibinfo{author}{\bibfnamefont{Y.~M.} \bibnamefont{Ivanchenko}}
  \bibnamefont{and} \bibinfo{author}{\bibfnamefont{T.~K.}
  \bibnamefont{Soboleva}}, \bibinfo{journal}{Phys.\ Lett.\ A}
  \textbf{\bibinfo{volume}{147}}, \bibinfo{pages}{65} (\bibinfo{year}{1990}).

\bibitem[{\citenamefont{Kogan}(1994)}]{Kogansemi}
\bibinfo{author}{\bibfnamefont{V.~G.} \bibnamefont{Kogan}},
  \bibinfo{journal}{Phys.\ Rev.\ B} \textbf{\bibinfo{volume}{49}},
  \bibinfo{pages}{15874} (\bibinfo{year}{1994}).

\bibitem[{\citenamefont{Thyssen et~al.}(1994)\citenamefont{Thyssen, Ustinov,
  Kohlstedt, Caputo, Pagano, and Flytzanis}}]{Caputo2}
\bibinfo{author}{\bibfnamefont{N.}~\bibnamefont{Thyssen}},
  \bibinfo{author}{\bibfnamefont{A.~V.} \bibnamefont{Ustinov}},
  \bibinfo{author}{\bibfnamefont{H.}~\bibnamefont{Kohlstedt}},
  \bibinfo{author}{\bibfnamefont{J.~G.} \bibnamefont{Caputo}},
  \bibinfo{author}{\bibfnamefont{S.}~\bibnamefont{Pagano}}, \bibnamefont{and}
  \bibinfo{author}{\bibfnamefont{N.}~\bibnamefont{Flytzanis}}, in
  \emph{\bibinfo{booktitle}{Proceedings of the International Conference on
  Nonlinear Superconducting Devices and High-Tc Materials}}, edited by
  \bibinfo{editor}{\bibfnamefont{R.~D.} \bibnamefont{Parmentier}}
  \bibnamefont{and} \bibinfo{editor}{\bibfnamefont{N.~F.}
  \bibnamefont{Pedersen}} (\bibinfo{publisher}{World Scientific, Singapore},
  \bibinfo{year}{1994}).

\bibitem[{\citenamefont{Ivanchenko}(1995)}]{Ivan}
\bibinfo{author}{\bibfnamefont{Y.~M.} \bibnamefont{Ivanchenko}},
  \bibinfo{journal}{Phys.\ Rev.\ B} \textbf{\bibinfo{volume}{52}},
  \bibinfo{pages}{79} (\bibinfo{year}{1995}).

\bibitem[{\citenamefont{Mints and Snapiro}(1995{\natexlab{a}})}]{MinSnap}
\bibinfo{author}{\bibfnamefont{R.~G.} \bibnamefont{Mints}} \bibnamefont{and}
  \bibinfo{author}{\bibfnamefont{I.~B.} \bibnamefont{Snapiro}},
  \bibinfo{journal}{Phys.\ Rev.\ B} \textbf{\bibinfo{volume}{51}},
  \bibinfo{pages}{3054} (\bibinfo{year}{1995}{\natexlab{a}}).

\bibitem[{\citenamefont{Mints and Snapiro}(1995{\natexlab{b}})}]{MinSnapcheren}
\bibinfo{author}{\bibfnamefont{R.~G.} \bibnamefont{Mints}} \bibnamefont{and}
  \bibinfo{author}{\bibfnamefont{I.~B.} \bibnamefont{Snapiro}},
  \bibinfo{journal}{Phys.\ Rev.\ B} \textbf{\bibinfo{volume}{52}},
  \bibinfo{pages}{9691} (\bibinfo{year}{1995}{\natexlab{b}}).

\bibitem[{\citenamefont{Kuzvovlev and Lomtev}(1997)}]{LomKuz}
\bibinfo{author}{\bibfnamefont{Y.~E.} \bibnamefont{Kuzvovlev}}
  \bibnamefont{and} \bibinfo{author}{\bibfnamefont{A.~I.}
  \bibnamefont{Lomtev}}, \bibinfo{journal}{JETP} \textbf{\bibinfo{volume}{84}},
  \bibinfo{pages}{986} (\bibinfo{year}{1997}).

\bibitem[{\citenamefont{Lomtev}(1998)}]{Lomtev}
\bibinfo{author}{\bibfnamefont{A.~I.} \bibnamefont{Lomtev}},
  \bibinfo{journal}{JETP} \textbf{\bibinfo{volume}{86}}, \bibinfo{pages}{1234}
  (\bibinfo{year}{1998}).

\bibitem[{\citenamefont{Caputo et~al.}(1999)\citenamefont{Caputo, Flytzanis,
  Kurin, Lazarides, and Vavalis}}]{Caputo1}
\bibinfo{author}{\bibfnamefont{J.~G.} \bibnamefont{Caputo}},
  \bibinfo{author}{\bibfnamefont{N.}~\bibnamefont{Flytzanis}},
  \bibinfo{author}{\bibfnamefont{V.~V.} \bibnamefont{Kurin}},
  \bibinfo{author}{\bibfnamefont{N.}~\bibnamefont{Lazarides}},
  \bibnamefont{and} \bibinfo{author}{\bibfnamefont{E.}~\bibnamefont{Vavalis}},
  \bibinfo{journal}{J. Appl. Phys.} \textbf{\bibinfo{volume}{85}},
  \bibinfo{pages}{7282} (\bibinfo{year}{1999}).

\bibitem[{\citenamefont{Flytzanis et~al.}(2000)\citenamefont{Flytzanis,
  Lazarides, Chiginev, Kurin, and Caputo}}]{Kurin}
\bibinfo{author}{\bibfnamefont{N.}~\bibnamefont{Flytzanis}},
  \bibinfo{author}{\bibfnamefont{N.}~\bibnamefont{Lazarides}},
  \bibinfo{author}{\bibfnamefont{A.}~\bibnamefont{Chiginev}},
  \bibinfo{author}{\bibfnamefont{V.~V.} \bibnamefont{Kurin}}, \bibnamefont{and}
  \bibinfo{author}{\bibfnamefont{J.~G.} \bibnamefont{Caputo}},
  \bibinfo{journal}{J. Appl. Phys} \textbf{\bibinfo{volume}{88}},
  \bibinfo{pages}{4201} (\bibinfo{year}{2000}).

\bibitem[{\citenamefont{Kogan et~al.}(2001)\citenamefont{Kogan, Dobrovitski,
  Clem, Mawatari, and Mints}}]{Kogan}
\bibinfo{author}{\bibfnamefont{V.~G.} \bibnamefont{Kogan}},
  \bibinfo{author}{\bibfnamefont{V.~V.} \bibnamefont{Dobrovitski}},
  \bibinfo{author}{\bibfnamefont{J.~R.} \bibnamefont{Clem}},
  \bibinfo{author}{\bibfnamefont{Y.}~\bibnamefont{Mawatari}}, \bibnamefont{and}
  \bibinfo{author}{\bibfnamefont{R.~G.} \bibnamefont{Mints}},
  \bibinfo{journal}{Phys.\ Rev.\ B} \textbf{\bibinfo{volume}{63}},
  \bibinfo{pages}{144501} (\bibinfo{year}{2001}).

\bibitem[{\citenamefont{Gr{\o}nbech-Jensen and Samuelsen}(2002)}]{groenbech}
\bibinfo{author}{\bibfnamefont{N.}~\bibnamefont{Gr{\o}nbech-Jensen}}
  \bibnamefont{and}
  \bibinfo{author}{\bibfnamefont{M.~R.}~\bibnamefont{Samuelsen}},
  \bibinfo{journal}{Phys.\ Rev.\ B} \textbf{\bibinfo{volume}{65}},
  \bibinfo{pages}{144512} (\bibinfo{year}{2002}).

\bibitem[{\citenamefont{Pearl}(1964)}]{Pearl}
\bibinfo{author}{\bibfnamefont{J.}~\bibnamefont{Pearl}},
  \bibinfo{journal}{Appl.\ Phys.\ Lett.} \textbf{\bibinfo{volume}{5}},
  \bibinfo{pages}{65} (\bibinfo{year}{1964}).

\bibitem[{\citenamefont{Dolan}(1977)}]{Dolan}
\bibinfo{author}{\bibfnamefont{G.~J.} \bibnamefont{Dolan}},
  \bibinfo{journal}{Appl. Phys. Lett.} \textbf{\bibinfo{volume}{31}},
  \bibinfo{pages}{337} (\bibinfo{year}{1977}).

\bibitem[{\citenamefont{Abramowitz and Stegun}(1965)}]{AbrSteg}
\bibinfo{editor}{\bibfnamefont{M.}~\bibnamefont{Abramowitz}} \bibnamefont{and}
  \bibinfo{editor}{\bibfnamefont{A.}~\bibnamefont{Stegun}}, eds.,
  \emph{\bibinfo{title}{Handbook of Mathematical Functions with Formulas,
  Graphs, and Mathematical Tables}} (\bibinfo{publisher}{U.S. GPO},
  \bibinfo{address}{Washington, D.C.}, \bibinfo{year}{1965}).

\bibitem[{\citenamefont{Hilgenkamp and Mannhart}(2002)}]{GrainRev}
\bibinfo{author}{\bibfnamefont{H.}~\bibnamefont{Hilgenkamp}} \bibnamefont{and}
  \bibinfo{author}{\bibfnamefont{J.}~\bibnamefont{Mannhart}},
  \bibinfo{journal}{Rev.\ Mod.\ Phys.} \textbf{\bibinfo{volume}{74}},
  \bibinfo{pages}{485} (\bibinfo{year}{2002}).

\bibitem[{\citenamefont{Kulik}(1965)}]{KulikFiske}
\bibinfo{author}{\bibfnamefont{I.~O.} \bibnamefont{Kulik}},
  \bibinfo{journal}{JETP Lett.} \textbf{\bibinfo{volume}{2}},
  \bibinfo{pages}{84} (\bibinfo{year}{1965}).

\bibitem[{\citenamefont{Kulik}(1967)}]{Kulik}
\bibinfo{author}{\bibfnamefont{I.~O.} \bibnamefont{Kulik}},
  \bibinfo{journal}{Sov.\ Phys.\ Tech.\ Phys.} \textbf{\bibinfo{volume}{12}},
  \bibinfo{pages}{111} (\bibinfo{year}{1967}).

\bibitem[{\citenamefont{Abdumalikov
  et~al.}(unpublished)\citenamefont{Abdumalikov, Jr., and Kurin}}]{Privat}
\bibinfo{author}{\bibfnamefont{A.~A.} \bibnamefont{Abdumalikov}},
  \bibinfo{author}{\bibnamefont{Jr.}}, \bibnamefont{and}
  \bibinfo{author}{\bibfnamefont{V.~V.} \bibnamefont{Kurin}}
  (\bibinfo{year}{unpublished}).

\bibitem[{\citenamefont{Lamb}(1980)}]{lamb}
\bibinfo{author}{\bibfnamefont{G.~L.} \bibnamefont{Lamb}},
  \emph{\bibinfo{title}{Elements of soliton theory}}
  (\bibinfo{publisher}{Wiley}, \bibinfo{address}{New York},
  \bibinfo{year}{1980}).

\bibitem[{\citenamefont{Martucciello and Monaco}(1996)}]{MonMart}
\bibinfo{author}{\bibfnamefont{N.}~\bibnamefont{Martucciello}} \bibnamefont{and}
  \bibinfo{author}{\bibfnamefont{R.}~\bibnamefont{Monaco}},
  \bibinfo{journal}{Phys. Rev. B} \textbf{\bibinfo{volume}{53}},
  \bibinfo{pages}{3471} (\bibinfo{year}{1996}).

\end{thebibliography}
\end{document}